\pdfoutput=1
\documentclass[a4paper,fleqn,usenatbib,useAMS]{mnras}

\usepackage{color}
\usepackage{aas_macros}
\usepackage{adjustbox}
\usepackage{bm} 
\usepackage{times}
\usepackage{amsmath}
\usepackage{cleveref}

\usepackage{graphicx}
\usepackage{amsmath}
\usepackage{breqn}
\usepackage{todonotes}
\usepackage{eso-pic}% http://ctan.org/pkg/eso-pic

\AddToShipoutPictureBG*{%
  \AtPageUpperLeft{%
    \hspace{0.75\paperwidth}%
    \raisebox{-5.5\baselineskip}{%
      \makebox[0pt][l]{\textnormal{DES 2015-0147}}
}}}%

\AddToShipoutPictureBG*{%
  \AtPageUpperLeft{%
    \hspace{0.75\paperwidth}%
    \raisebox{-6.5\baselineskip}{%
      \makebox[0pt][l]{\textnormal{FERMILAB-PUB-16-161-AE}}  
}}}%

\def\arcmin{\hbox{$^\prime$}}

\def\eprinttmp@#1arXiv:#2 [#3]#4@{
\ifthenelse{\equal{#3}{x}}{\href{http://arxiv.org/abs/#1}{#1}
}{\href{http://arxiv.org/abs/#2}{arXiv:#2} [#3]}}

\providecommand{\eprint}[1]{\eprinttmp@#1arXiv: [x]@}
\newcommand{\adsurl}[1]{\href{#1}{ADS}}

\title[Lognormality of DES galaxies and weak lensing convergence]{Testing the lognormality of the galaxy and weak lensing convergence distributions from Dark Energy Survey maps}

\author[]{
\parbox{\textwidth}{
\Large
L.~Clerkin$^{1}$,\thanks{Corresponding author: \texttt{lucyc@star.ucl.ac.uk}}
D.~Kirk$^{1}$,
M.~Manera$^{1}$,
O.~Lahav$^{1}$,
F.~Abdalla$^{1,2}$,
A.~Amara$^{3}$,
D.~Bacon$^{4}$,
C.~Chang$^{3}$,
E.~Gazta\~{n}aga$^{5}$,
A.~Hawken$^{6}$,
B.~Jain$^{7}$,
B.~Joachimi$^{1}$,
V.~Vikram$^{7}$,
T.~Abbott$^{8}$,
S.~Allam$^{9}$,
R.~Armstrong$^{10}$,
A.~Benoit-L{\'e}vy$^{11,1,12}$,
G.~M.~Bernstein$^{7}$,
R.~A.~Bernstein$^{13}$,
E.~Bertin$^{11,12}$,
D.~Brooks$^{1}$,
D.~L.~Burke$^{14,15}$,
A. Carnero Rosell$^{16,17}$,
M.~Carrasco~Kind$^{18,19}$,
M.~Crocce$^{5}$,
C.~E.~Cunha$^{14}$,
C.~B.~D'Andrea$^{4,20}$,
L.~N.~da Costa$^{16,17}$,
S.~Desai$^{21,22}$,
H.~T.~Diehl$^{9}$,
J.~P.~Dietrich$^{21,22}$,
T.~F.~Eifler$^{7,23}$,
A.~E.~Evrard$^{24,25}$,
B.~Flaugher$^{9}$,
P.~Fosalba$^{5}$,
J.~Frieman$^{9,26}$,
D.~W.~Gerdes$^{25}$,
D.~Gruen$^{14,15}$,
R.~A.~Gruendl$^{18,19}$,
G.~Gutierrez$^{9}$,
K.~Honscheid$^{27,28}$,
D.~J.~James$^{8}$,
S.~Kent$^{9}$,
K.~Kuehn$^{29}$,
N.~Kuropatkin$^{9}$,
M.~Lima$^{30,16}$,
P.~Melchior$^{10}$,
R.~Miquel$^{31,32}$,
B.~Nord$^{9}$,
A.~A.~Plazas$^{23}$,
A.~K.~Romer$^{33}$,
E.~Sanchez$^{34}$,
M.~Schubnell$^{24}$,
I.~Sevilla-Noarbe$^{34}$,
R.~C.~Smith$^{8}$,
M.~Soares-Santos$^{9}$,
F.~Sobreira$^{35,16}$,
E.~Suchyta$^{7}$,
M.~E.~C.~Swanson$^{19}$,
G.~Tarle$^{25}$,
A.~R.~Walker$^{8}$
 \emph{(Affiliations are listed at the end of the paper)}
}
}

\begin{document}

\date{}

\pagerange{\pageref{firstpage}--\pageref{lastpage}} 
\pubyear{2016}

\maketitle

\label{firstpage}

\begin{abstract}

It is well known that the probability distribution function (PDF) of galaxy density contrast is approximately lognormal; whether the PDF of mass fluctuations derived from weak lensing convergence ($\kappa_{\rm{WL}}$) is lognormal is less well established. We derive PDFs of the galaxy and projected matter density distributions via the Counts in Cells (CiC) method. We use maps of galaxies and weak lensing convergence produced from the Dark Energy Survey (DES) Science Verification data over 139 deg$^2$. We test whether the underlying density contrast is well described by a lognormal distribution for the galaxies, the convergence and their joint PDF. We confirm that the galaxy density contrast distribution is well modeled by a lognormal PDF convolved with Poisson noise at angular scales from 10\arcmin - 40\arcmin (corresponding to physical scales of 3--10 Mpc). We note that as $\kappa_{\rm{WL}}$ is a weighted sum of the mass fluctuations along the line of sight, its PDF is expected to be only approximately lognormal. We find that the $\kappa_{\rm{WL}}$ distribution is well modeled by a lognormal PDF convolved with Gaussian shape noise at scales between 10\arcmin and 20\arcmin, with a best-fit $\chi^2$/DOF of 1.11 compared to 1.84 for a Gaussian model, corresponding to p-values 0.35 and 0.07 respectively, at a scale of 10\arcmin. Above 20\arcmin a simple Gaussian model is sufficient. The joint PDF is also reasonably fitted by a bivariate lognormal. As a consistency check we compare the variances derived from the lognormal modelling with those directly measured via CiC. Our methods are validated against maps from the MICE Grand Challenge N-body simulation. 

\end{abstract}

%\begin{keywords}
%%gravitational lensing: weak; methods: data analysis; cosmic microwave background\vspace{-1.7cm}
%\end{keywords}

%\newpage
%\clearpage
%----------------------------------------------------------------------------------------

%-----------------------------------------------------

\section{Introduction}
\label{sec:intro}

It was first noted by Hubble that the distribution of galaxies in angular cells on the celestial sphere is well approximated by a lognormal \citep{Hubble1934}. This has been confirmed observationally (\citealt*{Coles1991}, \citealt{Wild2005}) as well as in N-body simulations (\citealt*{Bernardeau_Kofman_1995}, \citealt{Bernardeau1994}, \citealt*{Kayo2001}), which have shown that the underlying mass density field is expected to be lognormal.

Since the weak lensing convergence field along the line of sight is a weighted projection of the mass density contrast field, one might suspect that the lognormal distribution is a reasonable, if not exact, model of this too. This has been tested on numerical simulations and a lognormal PDF shown to be a reasonable model (\citealt*{Taruya2002}, \citealt*{Hilbert2011a}). Even better fits to the convergence PDF, particularly in the tails of the distribution, have been obtained by generalisations of a lognormal PDF (\citealt*{Das2006}, \citealt{Takahashi2011}, \citealt*{Joachimi2011a}). 

The Dark Energy Survey (DES) \citep{DES2005,DES2015,DES2016} presents an excellent opportunity to study both of these fields. DES was specifically conceived to produce cutting edge science from four different cosmological probes - large-scale structure, weak gravitational lensing, galaxy clusters and supernovae - using the same instrument. The full survey involves five years of observations, currently in progress. In this paper we focus on data produced during the pre-survey Science Verification (SV) series of observations. 

This early data from DES allowed for the construction of two types of density fields. One is from luminous matter, i.e. galaxies of various types, $\delta_g$, which are biased tracers of the underlying dark matter field, $\delta_m$. The other uses the weak lensing of galaxy shapes to construct a convergence, or $\kappa$ map \citep{Vikram2015, Chang2015} that is directly sensitive to the integrated dark matter field out to the lensed galaxies.

Both maps trace the underlying density distribution in the Universe. Galaxies are biased tracers of matter density, preferentially clustering in overdense regions. Galaxy density contrast can then be considered a biased local tracer of the density field.

Weak lensing convergence on the other hand responds directly to the underlying density field and is therefore unbiased. However gravitational deflection of light is a cumulative effect, sensitive to the integrated matter density along the line of sight from source galaxy to observer. The convergence field for a given galaxy source distribution therefore gives us information about the cumulative density field between observer and source, with the exact contribution of matter at different distances along the line of sight governed by the lensing efficiency function.

%The CFHTLenS mass-mapping paper of [REF VanW] presented measurements of moments of the WL convergence distribution as well as a 1D PDF based on smoothed simulated maps. 

The purpose of the present study is to analyse the galaxy and mass maps from DES SV simultaneously, testing the two maps separately for log-normality, as well as analysing the joint distribution. To our knowledge this is the first time that the log-normality of the weak lensing convergence field alone has been tested using data rather than numerical simulation \citep{Taruya2002}, and the first time the joint distribution has been tested for log-normality. 

The \emph{Counts in Cells} (CiC) method (e.g. \citealt{Hubble1934, White1979, Gaztanaga1992, Szapudi1997,Bernardeau2002}) is a natural way to measure the individual and joint PDFs. The CiC technique splits up a particular data set into spatial cells, in two or three dimensions, and takes an aggregate of the available information inside each cell. 
% The simplest case is galaxy number counts where we simply count the number of galaxies inside each cell but the technique can be employed more generally. 
Statistical variation between cells then provides information on the properties of that cosmological field. 
DES observations are ideally suited to this sort of analysis. The fact that DES provides a joint galaxy survey and convergence map data set produced from the same observations makes it easier to ensure consistency between data and to control for systematics. The SV data we use were taken before the start of the full five year survey, covers 139 deg$^2$ to full survey depth and forms a test-bed for the kind of analyses planned on the final DES data. All of the analyses in this work are done first on galaxy and convergence maps from MICE simulations in order to test our methodology.

% The analysis is done using weak lensing convergence rather than shear for two reasons: firstly because convergence is a scalar quantity meaning the analysis is more straightforwardly carried out, and secondly because the DES convergence map has been produced (Vikram et al. 2015, Chang et al. 2015).

% To optimise the correlation between our galaxy and WL convergence field we use the $\kappa_{\rm gal}$ re-weighting proposed by \citet{amara2012} and performed in \citet{VCJ+15}. This re-weighting generates the ``convergence'' map which would be seen if the matter density field responsible for the gravitational lensing corresponding exactly to the measured galaxy over-density field. %Using the bivariate log-normal fit to our convergence and galaxy-convergence maps we measure the relative stochasticity and non-linearity between the two fields.

% Our goals with this work are to test to what extent the DES SV galaxy and convergence fields can be described as lognormal, to exploit the joint moments of the galaxy and convergence maps to learn about galaxy biasing and to use the fact that CiC accesses the full PDF of a cosmic field to quantify non-linear and stochastic clustering in the DES SV data through measurement of skewness and kurtosis.
The outline of the paper is as follows. In section \ref{sec:LN_theory} we review the theory and formalism used. We describe the galaxy and weak lensing convergence maps from MICE simulations and DES used in section \ref{sec:data}, and summarise our CiC method in section \ref{sec:method}. In section \ref{sec:MICE} we validate our CiC method on MICE Grand Challenge N-body simulations, checking that we see the expected lognormality in MICE $\delta_g$ and the noise-free convergence. In section \ref{sec:LN_DES} we present lognormal fits to the individual DES galaxy and convergence field distributions as well as their joint distribution. We check the validity of the log-normal model by measuring the variance of the fields and comparing this to the variance derived under the assumption of log-normality. We discuss the results in \ref{sec:conc}, and in the Appendices we give the formalism used to calculate moments from CiC, test the impact of systematic effects, and confirm that assumptions we make in our method do not significantly affect our results.

%In \ref{sec:NL_stoch} we use the properties of the log-normal model to quantify the non-linear and stochastic bias between the galaxy and convergence fields,

%We use two appendices to present systematic tests of our method and data. In \ref{sec:ApA} we concentrate on observational and astrophysical systematics in the data, while in \ref{sec:ApB} we confirm that assumptions we have made in the implementation of our method do not significantly effect our results.

%%%%%%%%%%%%%%%%%%%%%%%%%%%%%%%%%%%%%%%%%%%%%%%%%%%%%%%%%%%%%%%%%%%%%%%%%%%%%%%%%%%%%%%%%%%%%%%%%%%%%%%%%%%%%%%

\section{Lognormal Modelling of cosmic fields}
\label{sec:LN_theory}

% Cosmological density fields are often characterised as Gaussian random fields at linear order but, as gravitational instability leads to the growth of nonlinear structures, this cannot be a complete description. One stochastic model that has found favour in the modelling of galaxy distributions is the lognormal (LN) random field \citep{CJ91}. This is a simple model but fits the purpose well, as it ensures $\rho > 0$, and approaches the Gaussian distribution at early times.
% \subsection{Lognormal Distributions in Nature}
Lognormal distributions are very common in nature, from the sizes of clouds, pebbles on a beach, or crystals in icecream; the length of sentences or words in a conversation; to populations of bacteria (see \citealt*{Limpert2001}, \citealt*{Gaskell2004} and references therein). Many of these examples involve multiplicative processes, of either merging or fragmentation. Any process that can be written as a product of terms will, if there are many terms, tend to a lognormal distribution. This is because if a process $X$ can be written as a product of independent terms, then $\rm{ln}(X)$ will be a sum of independent terms, and via the central limit theorem these will be normally distributed. So $\rm{ln}(X)$ is normally distributed, or, $X$ is lognormally distributed. 
%One can think of the lognormal as applying to non-linear processes in the way that the normal distribution applies to linear ones.

There are many examples of the hierarchical merging or fragmenting of structure leading to lognormal distributions, such as: the initial mass function of field stars, explained in terms of cloud fragmentation \citep{Zinnecker1984}; X-ray flux variations, suggesting lognormal distribution of emitting regions \citep{Gaskell2004}; luminosity functions of central galaxies, explained in terms of BCGs being formed by steps of mass adding/stripping (e.g. \citealt*{Taghizadeh-Popp2012}); and the angular momentum of disc galaxies \citep{Marr2015a}. 

In this paper we test the lognormality of the distribution of matter in the Universe via the galaxy density contrast field, $\delta_g$, and via the weak lensing convergence field, $\kappa_{\rm{WL}}$. Each approach has particular observational and astrophysical noise associated with it, which we discuss and propose models for in this section.

\subsection{Galaxies}

% \cite{Kayo2001} showed that, for N-body simulations, a log-normal distribution well approximated the nonlinear mass distribution, evolved from Gaussian initial conditions, for both one- and two-point statistics. This result held regardless of the shape of the initial power spectrum. Wild et al (2005) confirmed that the joint distribution of blue and red 2dF galaxies in the 2dF Galaxy Redshift survey is well approximated by a log normal.
In the standard picture of gravitational instability, the primordial density fluctuations that are the precursor of all structure in the universe are assumed to be a random Gaussian field. Once they enter the non-linear regime, with finite rms fluctuations, their PDF must deviate from a Gaussian to avoid non-zero probabilities being attached to regions with negative densities \citep{Fry1984}. The exact form of the PDFs in this regime is not known but there are various phenomenological models that are fully specified statistically and satisfy the common sense requirement that the matter density, $\rho \geqslant 0$ (e.g. \citealt*{Saslaw1984}, \citealt*{Suto1990}, \citealt{Lahav1993}, \citealt{Gaztanaga1993a}; \citealt{Suto1993}, \citealt*{Ueda1996}). 

One such model that is widely used is the lognormal. As well as being completely specified statistically and always having $\rho \geqslant 0$, it becomes arbitrarily close to Gaussian statistics at early times and has the advantage that it can be handled analytically.  The merits of this model in a cosmological context are discussed extensively in \citet{Coles1991}. They explain possible motivations for using the lognormal model as: empirical; kinematic; an application of the central limit theorem (as described above); and importantly, simplicity. It is one of the few random fields for which interesting quantities such as its moments can be calculated analytically. 

It should be noted that despite these compelling motivations to use a lognormal in the statistical treatment of density perturbations, it does have shortcomings. In particular, it is not uniquely specified by its moments; many distributions can lead to the same set of moments. It must then be the case that information is lost in going from a lognormal field to its moments, an effect quantified in \citet{Carron2011}. However, it remains a popular and useful tool in analysis of the mass density contrast field.

Galaxies are biased tracers of the mass density contrast field. The 1D log-normal distribution of galaxy density contrast $\delta_g$ = $(\rho - \bar\rho)/\bar\rho$ is given by:
\begin{equation} 
	f(\delta_g) \mathrm{d}\delta_g = \frac{1}{w \sqrt{2 \pi}}exp\left( \frac{-x^2}{2w^2}\right) \mathrm{d}x
	\label{eqn:LN}
\end{equation}
where $x = \mathrm{ln} (1+\delta_g) + w^2/2$ and $w^2$ is the variance of the corresponding normal distribution $f[ln(1+\delta_g)]$. The offset $w^2/2$ ensures that $\langle \delta_g \rangle = 0$. The width $w$ is then the single free parameter of the 1D lognormal distribution.

If the lognormal distribution correctly describes the data, the variance of the overdensities will be related to the variance, $w$, of the underlying Gaussian distribution by 
\begin{equation}\label{eq:gg}
\langle\delta_g\delta_g\rangle = \mathrm{e}^{w^2}-1.
\end{equation}
Due to the discrete nature of galaxies, shot noise is present. Assuming Poisson sampling of galaxies,  %The use of galaxy shears to construct the $\kappa_{WL}$ suffers from similar noise and an additional shape noise term due to the width of the distribution of intrinsic galaxy ellipticities. The wider this intrinsic distribution the greater the shot noise contribution.
the shot noise in the measurement of the distribution of galaxy overdensities can be accounted for by convolving the log-normal model with a Poisson distribution. The probability distribution function of the galaxy counts $N$ in a cell of given size is then given by:

\begin{equation}\label{eq:PxLN}
	P(N) = \int^\infty_{-1} \frac{\bar{N}^N (1+\delta_g)^N}{N!} e^ {-\bar{N}(1+\delta_g)} f(\delta_g) d\delta_g 
\end{equation}
This Poisson sampled lognormal distribution has been shown to be a good fit to different galaxy populations in \citet{Coles1991}, \citet{Blanton2000a} and \citet{Wild2005}. In this work the smallest number of DES galaxies in a cell  considered is around 300, so the shot noise term is important.

\subsection{Weak Lensing Convergence}\label{sec:LN_theory:WL}

Various expressions for the convergence PDF have been proposed \citep{Munshi2000a,Valageas2000,Kainulainen2011a}. The lognormal model has the advantage - as in the case of the matter density contrast - of mathematical convenience, while offering the chance to extract more information than assuming a purely Gaussian model for the convergence field. Following a study that showed that a lognormal transformation of the matter density contrast increases the signal to noise \citep{Neyrinck2009}, \citet{Seo2011} performed an analogous study of the weak lensing convergence. They found that such a transform, when applied to the positively offset convergence, decorrelated angular frequencies and increased the signal-to-noise in the transformed power spectrum.

The convergence field along a line-of-sight can be expressed as a weighted projection of the mass density contrast field:

\begin{equation}
\kappa(\theta) = \int^{\chi_{\rm hor}}_{0} d\chi w(\chi)\delta[r(\chi)\theta,\chi], 
\label{eqn:kappa_from_delta}
\end{equation}
where $\chi$ is the comoving distance, $\chi_{\rm hor}$ is the angular diameter distance to the horizon and $\delta[r(\chi)\theta,\chi]$ is the underlying matter density contrast field. $w(\chi)$ is a geometrical weight function that depends on the relative separations of sources, lens and the observer (see e.g. \citealt{Mellier1999}; \citealt{Bartelmann2001} for reviews). It takes the form 

\begin{equation}
w(\chi) = \frac{3}{2}\left(\frac{H_{0}}{c}\right)^{2} \frac{\chi\Omega_{0}}{a(\chi)} \int_{\chi}^{\chi_{\rm hor}} d\chi' n(\chi')  \frac{\chi'-\chi}{\chi'} , 
\label{eqn:lensing_weight_fn}
\end{equation}
where $n(\chi)$ is the source galaxy distribution.

The distribution of $\kappa$ is not expected to be exactly lognormal, even if $\delta$ is, since $\kappa$ is a weighted projection of the mass density contrast field along line of sight. However, simulations have shown \citep{Taruya2002} that the convergence field is well approximated by a lognormal outside the regime of extremely high $\kappa$. Hence we choose in this work to model the noise free $\kappa$ field distribution with a shifted lognormal
% To model the convergence with a log-normal distribution we introduce the normalised convergence field, $\hat{\kappa}\equiv \kappa/|\kappa_{min}|$ (\cite{Taruya2002}), which imposes the required range $0 < \hat{\kappa} < \infty$. We can then apply the log normal model to the convergence field in the same way as the galaxy overdensity field, but with the substitutions $\delta \rightarrow \hat{\kappa}$. This gives an analogue of equation \ref{eqn:LN} for the convergence field,

\begin{equation}\label{eq:WL_LN}
P(\kappa) = \begin{cases} 
	\cfrac{\exp\Biggl[-
    \cfrac{(\textnormal{ln}(\kappa_0+\kappa)-\mu)^2}{2\sigma^2}\Biggr] }
    {\textnormal{ln}(\kappa_0+\kappa)  \sqrt{2 \pi}\sigma }
    & \text{for } \kappa > - \kappa_0,\\
    0 & \text{otherwise},
	\end{cases}
\end{equation}
where the shift $\kappa_0 = -\kappa_{min}$ and is called the `minimum convergence parameter'  \citep{Hilbert2011a}. The lowest possible value of $\kappa$ is given by $-\kappa_0$ since the lognormal is only defined for a positive range. The mean is given by 

\begin{equation}\label{eq:mu}
\mu = \textnormal{ln}(\kappa_0+\langle\kappa\rangle) - \sigma^2 / 2
\end{equation}
and the second moment
\begin{equation}\label{eq:kk}
\langle\kappa\kappa\rangle = e^{(2\mu+\sigma^2)}(e^{\sigma^2}-1).
\end{equation}
The value assigned to $\kappa_0$ is a modelling choice that can be approached in several ways. The minimum measured value of $\kappa$ could be used, but this is a noisy quantity and should not be used unless one has access to many realisation of $\kappa$. Or, treating $\kappa$ analogously to $\delta_g$, we could consider a theoretical minimum corresponding to the convergence we would see, for a given source distribution, if the mass distribution was a pure void along the entire line of sight. For the MICE source distribution used in this work this value is $-0.050$, and for the DES source distribution it is $-0.053$. However simulations show that there are no empty lines of sight in a $\Lambda$CDM universe \citep{Taruya2002, Vale2003, Hilbert2011a}. So we choose, where possible, to treat $\kappa_0$ as a free parameter and fit it jointly with the lognormal width.  

% One can think of equation \ref{eq:WL_LN} as an analogue to the lognormal model of the galaxy overdensity field, equation \ref{eqn:LN}, where the shift is the minimum value of $\delta$ theoretically possible, i.e. in a pure void along the line of sight, where $\delta = -1$. $\kappa_0$ can be fitted as a free parameter, or  A `theoretical minimum' for $\kappa$ can similarly be calculated as the convergence we would see, for a given source distribution, if the mass distribution was a pure void along the entire line of sight.  

% [** INSERT COMMENT ABOUT ABILITY TO JOINTLY CONSTRAIN SIGMA AND K0, AND WHEN THEY ARE DEGENERATE?**]

\begin{figure}
		\centering
        \includegraphics[scale=0.4]{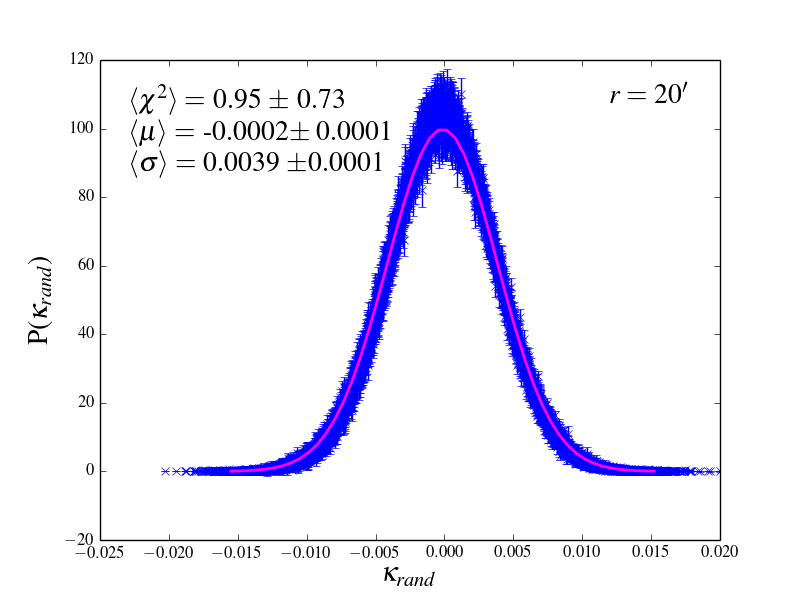}
		\caption{\label{fig:100_Krand} Demonstration of the Gaussianity of the noise in DES weak lensing convergence, $\kappa$, at a smoothing scale of 20\arcmin. Probability density distributions of the 100 realisations of the $\kappa$ map, in which shears were randomised, are shown in blue, with jackknife errors. A Gaussian PDF was fitted to each, with the mean of the best fitting PDFs shown in magenta. A Gaussian model is an excellent fit to the noise, with average goodness of fit $\chi^2/$DOF$ = 0.95 \pm 0.73$. }
\end{figure}

As for galaxies, we need to modify the lognormal to account for noise. The DES $\kappa$ map is constructed (see section \ref{data:DES_K}) from measurements of shear, which is the change  in  the  ellipticity  of  galaxies  resulting  from  weak  gravitational  lensing. Since galaxies are intrinsically elliptical (i.e. in the absence of lensing), the observed shear is the sum of this intrinsic ellipticity and the shear caused by lensing. The variance of the intrinsic ellipticity, called shape noise, is the dominant source of noise in shear measurements, typically by a factor of more than an order of magnitude. An estimate of the shape noise in the DES $\kappa$ map is provided by the 100 noise realisations described in section \ref{data:DES_K}. 

To analyse the shape of the noise distribution we construct PDFs via CiC (as described in section \ref{sec:method}) on each of the 100 maps. The resulting distributions appear Gaussian, as shown in figure \ref{fig:100_Krand}, where the thick blue curve is made up of 100 superimposed noise distributions with jackknife error bars, and the magenta line shows the mean bestfit Gaussian PDF. A Gaussian model provides an excellent fit, with average goodness of fit over the 100 maps $\chi^2/dof = 0.95 \pm 0.73$. %All goodness of fit values quoted hereafter are $\chi^2$\DOF, i.e. $\chi^2$ per degree of freedom. 

% ** Add sentence or two about B-mode **

We therefore propose that the 1D probability distribution for the weak lensing convergence field is then given by a convolution of a lognormal distribution with a noise contribution modelled as Gaussian:

\begin{equation}\label{eq:GxLN}
	f(\kappa) = \frac{1}{\sqrt{2 \pi} \sigma_n} \int^\infty_{-\kappa_0} \exp\Big[-\frac{(\kappa'-\kappa)^2}{2\sigma_n^2}\Big]  P(\kappa') \mathrm{d}\kappa'
\end{equation}
where $ P(\kappa)$ is the noise free log-normal distribution given in equation \ref{eq:WL_LN}, and $\sigma_n$ is the Gaussian width of the shape noise. In this work, $\sigma_n$ is determined from the 100 noise realisations.

\subsection{Joint Galaxy and Weak Lensing Convergence Field}

We can try to model the joint distribution of galaxy density contrast $\delta_{g}$ and weak lensing convergence $\kappa_{\rm{WL}}$ as a bivariate lognormal with PDF $f(\delta_{g},\kappa_{\rm{WL}})$. Following the notation used in \citet{Coles1991,Wild2005}, this is given by

\begin{equation}
f(\delta_g,\kappa) = \frac{|V|^{-1/2}}{2\pi} \exp \left[ -\frac{(\tilde{g}_{\delta}^{2} + \tilde{g}_{\kappa}^{2} - 2r_{\text{LN}}\tilde{g}_{\delta}\tilde{g}_{\kappa})}{2(1-r^{2}_{\text{LN}})} \right],
\label{eqn:2D_LN}
\end{equation}
where $g_x = \ln(x) - \left<\ln(x)\right>$, with $x = (1+\delta_g)$ and $x = (1+\kappa/\kappa_0)$ for the galaxy and convergence fields respectively, and $\tilde{g_x} = g_x/\omega_x$ where $\omega_x$ is the variance of the underlying Gaussian field $\ln(x)$. 

The lognormal correlation coefficient $r_{\text{LN}}$ is given by
\begin{equation}
r_{\text{LN}} = \frac{\left< g_{\delta}g_{\kappa} \right>}{\omega_{\delta}\omega_{\kappa}} \equiv \frac{\omega^{2}_{\delta\kappa}}{\omega_{\delta}\omega_{\kappa}}
\end{equation}
and $|V|$ is the determinant of the covariance matrix
\begin{equation}
V = \left( \begin{array}{cc}
\omega^{2}_{\delta} & \omega^{2}_{\delta\kappa} \\
\omega^{2}_{\delta\kappa} & \omega^{2}_{\kappa} \end{array} \right).
\end{equation}
Note that $r_{\text{LN}}$ and $V$ are defined in log-density space, and so $r_{\rm{LN}}$ is not the same as the (linear) Pearson correlation coefficient $\rho$.
The conditional probability
\begin{eqnarray}\label{eq:cond}
% \begin{equation}
f(\delta_{g}|\kappa) &=& \frac{f(\delta_{g},\kappa)}{f(\delta_{g})}\\
    &=& \frac{w_{\delta}}{(2 \pi |V|)^{1/2}} 
    exp\Biggl[-\frac{(\tilde{g}_{\delta} - r_{\rm{LN}}\tilde{g}_{\kappa})^2}
    {2(1-r^{2}_{\rm{LN}})}\Biggr].
\end{eqnarray}
Since $f(\delta_g,\kappa) = f(\kappa)f(\delta_g|\kappa)$ we can combine equations \ref{eq:GxLN} and \ref{eq:cond} to give the joint probability distribution function $f(\delta_g,\kappa)$. Including the convolutions with Poisson shot noise and Gaussian shape noise then gives

\begin{multline}\label{eq:biv}
P(N,\kappa) =  \int^\infty_{-1} \int^\infty_{-\kappa_0} 
\frac{1}{\sqrt{2 \pi} \sigma_n} \exp\Big[-\frac{(\kappa'-\kappa)^2}{2\sigma_n^2}\Big]  f(\kappa) \\
\times \frac{\bar{N}^N (1+\delta_g)^N}{N!} e^ {-\bar{N}(1+\delta_g)} f(\delta_g|\kappa)\mathrm{d}\delta_g \mathrm{d}\kappa'
\end{multline}

% \end{equation}

\section{The Data}
\label{sec:data}

\begin{figure}
		\centering
		\includegraphics[scale=0.2]{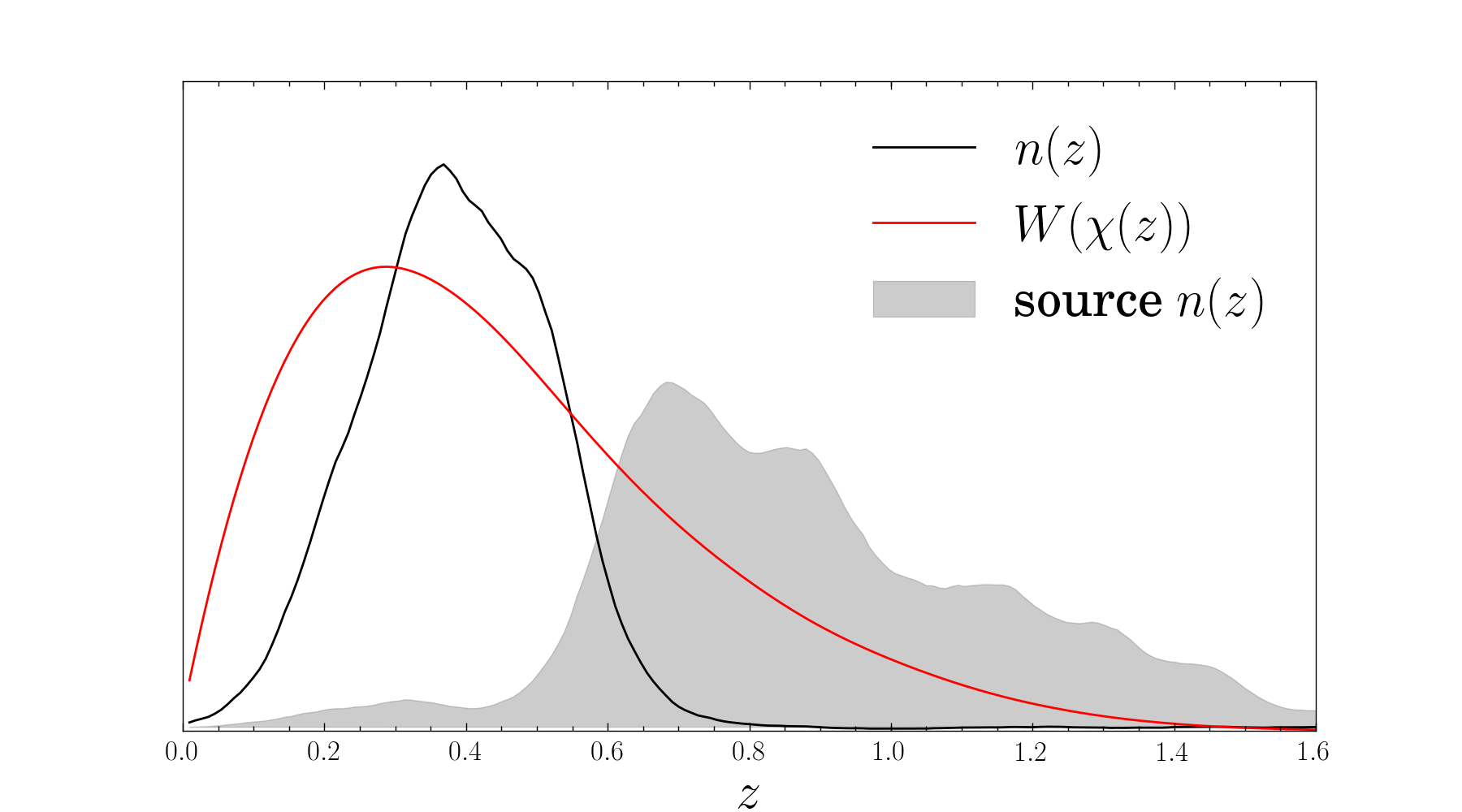}
		\caption{\label{fig:z_kernels} Redshift kernels of the observables considered in this paper: the galaxy redshift distribution of the DES Benchmark galaxy sample using the best-fit Skynet photo-z estimation (black line), and the lensing efficiency function of the sources used to make the DES $\kappa$ map (red line). Also shown is the redshift distribution of the source galaxies (shaded region). Each is shown with an arbitrary normalisation to make comparison easier.}
\end{figure}

This paper uses the DES Science Verfication (SV) galaxy and shape catalogues. The SV data were gathered between November 2012 and February 2013, shortly after DECam \citep{Flaugher2015} commissioning, and before the beginning of the (five year) DES survey proper in August 2013. The operation of the camera, survey planning, data analysis and reduction were all tested in preparation for starting year one of DES itself. The SV goal was to reproduce the properties of the full five-year DES survey over a much smaller sky area. 
%In practice weather and teething issues lead to some complications, see \citet{leistedtea2015} for more details.

Five optical filters ($grizY$) are used, with exposure times of 90 seconds for $griz$ and 45 seconds for $Y$. The final median depth in our region of interest, per band, was $g \sim 24.0$, $r \sim 23.9$, $i \sim 23.0$ and $z \sim 22.3$.

In total the SV data covered $\sim 250$ deg$^2$ at close to the nominal depth of the full DES survey. The observing footprint was divided into regions to maximise overlap with other surveys and with several small fields used for SNe searches.

In this paper we concentrate on a large contiguous region of $\sim$139 deg$^2$ called the SPT-E field due to its overlap with the South Pole Telescope CMB survey. This amount of contiguous data makes the SV SPT-E field a powerful data set in its own right, particularly for weak gravitational lensing where it rivals the full CFHTLenS \citep{Erben2013} and is only slightly shallower. 

% More detail on the SV observations, data reduction and the production of the \emph{SVA1 Gold} and bench catalogues can be found in [REF: Crocce et al.]. 

% \subsection{DES SV Data}
% \label{data:DES}

\subsection{DES Galaxy Sample}
\label{data:DES:DES_gals}

We use a particular subset of the DES SV galaxy catalogue known as the ``Benchmark'' sample \citep{Crocce2015}. First a catalogue of galaxies suitable for LSS analysis was constructed from the SV data and dubbed the ``Gold'' sample \citep{Rykoff2016}. Objects were included if detected in all five of the DES photometric bands. This covered $\sim$244 deg$^2$, restricted to $\textit{dec} > -61$ to avoid the Large Magellenic Cloud and R Doradus regions. In addition the Gold catalogue included masking of satellite trails and other artifacts, removal of regions where colors are severely affected by stray light and the application of additional stellar locus correction \citep{Kelly2014}.

From this Gold sample, the Benchmark sample was selected for cosmological analysis by imposing the additional conditions:

\begin{itemize}
\item $18.0 < i < 22.5$ 
\item $0 < g-r < 3$, $0 < r-i < 2$ and $0 < i-z < 3$.
\item \texttt{wavg\_spread\_model} $>$ 0.003 (star-galaxy separation) 
\item $60 < ra < 95$ and $-62 < dec < -40$ (SPT-E),
\end{itemize}
where $i$ refers to \texttt{SExtractor}'s \texttt{MAG_AUTO} quantity. The cuts on position restrict our analysis to the SPT-E region. The redshifts used in this paper come from the Skynet photo-z pipeline (\citealt{Bonnett2015a}, \citealt{Graff2013}). 
%For the galaxy sample used throughout this paper we remove galaxies with a mean redshift outside the range $0.6<z<1.3$ because the pipeline was shown to be inaccurate for sources outside this range in a stringent series of tests optimised for WL analysis \citep{bonnettetal2015a}. 
The galaxy redshift distribution is shown in figure \ref{fig:z_kernels}. The redshift range we use throughout this paper is $0.1<z<1.5$, chosen as in this region the galaxy redshift distribution overlaps with the lensing efficiency function used to make the DES $\kappa_{WL}$ map (see next section).

% \dk{Update this paragraph when we have confirmed final redshifts we plan to use.}
%The Benchmark catalogue has some particular properties that are useful for our analysis. It is complete down to $i=22.5$ over the redshift range in question, see Fig. \ref{fig:des_gal_mag_z}. This allows us to perform our CiC analysis without needing to account for a complex selection function. The \texttt{modest\_class} requirement refers to a star/galaxy classification scheme employed within the collaboration and discussed in more detail in [REF: Crocce, desai et al 2012, Soumagnac et al 2013]. 
% and the benchmark SPT-E footprint is shown in Fig. \ref{fig:sv_map}

% \begin{figure}
% 		\centering
% 		\includegraphics[scale=0.4]{abs_mag_i_vs_z_tpz_bench.png}
% 		\caption{\label{fig:des_gal_mag_z} Plot of i-band absolute magnitude vs. redshift for the DES Bench galaxy sample. The absolute magnitudes are calculated by [detail] and the redshifts used are the best-fit $z$-estimates from the TPZ photo-z pipeline.}
% \end{figure}

% \begin{figure}
% 		\centering
% 		\includegraphics[scale=0.4]{sv_map.png}
% 		\caption{\label{fig:sv_map} Map of the DES galaxy overdensity used in our survey. Regions that were unobserved or masked are in white. The map is pixellated on a HEALPix grid of $N_{side}=4096$.}
% \end{figure}

\subsection{DES $\kappa$ Map}
\label{data:DES_K}

Shear measurement on DES SV galaxy images was performed with two independent pipelines: \textsc{im3shape}\footnote{The open source code can be downloaded at: \texttt{https://bitbucket.org/joezuntz/im3shape/}} \citep{Zuntz2013} and \textsc{ngmix}\footnote{The open source code can be downloaded at: \texttt{https://github.com/esheldon/ngmix}} \citep{Sheldon2014}. %The maps in this paper are constructed from the version 9 and version 011 catalogues respectively.  

Extensive testing of both codes was carried out by the DES collaboration (see \citealt{Jarvis2015} for details) and both pipelines passed all requirement tests for measurement of cosmic shear with the SV data set. A number of cuts were applied to both catalogues to remove stars, spurious detections, poor measurements and other effects that could bias shear measurement; these are also described in \cite{Jarvis2015}. %For the maps used in this paper the ``conservative additive'' selection was applied which acts to remove galaxies with S/N<20 from \texttt{ngmix} and S/N$<$15 from \texttt{im3shape}, as well as very small galaxies. 

Shear measurements for a given galaxy are headless vectors and the cosmic shear field is therefore a spin-2 quantity. To allow us to perform our CiC analysis on a scalar quantity we work with maps of weak lensing convergence, $\kappa$, a spin-0 field. This $\kappa$-reconstruction was performed using the Kaiser-Squires method \citep{Kaiser1993}, and the production and initial analysis of these $\kappa$ maps is described in detail in \citet{Vikram2015}.

The Kaiser-Squires reconstruction method uses the relation of the Fourier transform of the observed shear, $\hat{\gamma}$, to that of the convergence, $\hat{\kappa}$,
\begin{equation}
\hat{\kappa}_{\ell} = D^*_{\ell}\hat{\gamma}_\ell,
\label{eqn:FT}
\end{equation}
\begin{equation}
D_{\ell} = \frac{\ell_{1}^{2}-\ell_{2}^{2}+2i\ell_{1}\ell_{2}}{|\ell|^{2}},
\end{equation}
where $\ell_i$ are the Fourier counterparts of the angular coordinates, $\theta_i, i=1,2$. The inverse Fourier transform of equation \ref{eqn:FT} gives the convergence for the observed field in real space. In the absence of noise, systematics and masking, the convergence will be a real (spin-0) quantity. In reality these effects produce a non-zero imaginary component. It is most convenient to express the real part of the convergence map as a map of curl free E-modes, and the imaginary part as divergence free B-modes. The $\kappa$ maps have pixels of size 2\arcmin. For use in this analysis the original flat sky $\kappa$ maps are transformed into \texttt{HEALPix} \citep{Gorski2005} maps at resolution $N_{\text{side}}$=4096. This is done by dividing each pixel of the flat sky maps into 25 sub-pixels, and creating a \texttt{HEALPix} map by combining these sub-pixels. This procedure reduces inaccuracies in changing from one mapping system to another, and in tests gives the same angular power spectrum measurements as the flat sky map to well within the errors.

The source galaxy selection used to construct the $\kappa$ map used in this paper took galaxies with redshifts in the range $0.6 < z < 1.3$. The resulting redshift efficiency function is shown in figure \ref{fig:z_kernels}. The lensing efficiency function peaks at {$z \sim 0.3$}, and our selection of galaxies at $0.1 < z < 0.5$ overlaps significantly with the range of redshifts to which the $\kappa$ map is sensitive.
%as such most of the galaxies in the upper half of our galaxy redshift distribution do not overlap significantly with the range of redshifts to which the mass maps are sensitive. 
%For this reason we construct an additional set of galaxy overdensity maps with a maximum redshift of $z<0.7$ and use this sample for all cross-correlations between $\delta_{\rm g}$ and $\kappa$. 

%Unlike the galaxy sample there is no need for the weak lensing source sample to be complete. It is more important that we select galaxies with accurate shear measurements because the observed shear signal is sensitive to all the mass between the source and the observer. 
% \dk{More detail here on cuts in shear catalogues.}

In addition to the E-mode $\kappa$ map we make use of a number of other products made in the course of the DES mass-mapping analysis. A B-mode map was constructed by rotating the measured galaxy ellipticities by 45 degrees. The physical process of weak gravitational lensing does not induce B-modes in the convergence field so the B-mode map is a test of systematic effects in our observations, shear measurement and $\kappa$-reconstruction; it should be consistent with zero within our reconstruction noise. We will refer to the B-mode reconstructed map as $\kappa_B$. 

In addition to the E- and B-mode maps we also make use of a series of noise-only realisations, made by taking the galaxy shape catalogue and rotating the measured shape of each galaxy by some random angle. $\kappa$ maps were then constructed from each randomised catalogue in the usual way. This has the effect of destroying all cosmological information in the resulting maps, while retaining the same noise properties as the data (because the distribution of galaxies on the sky and in redshift remains the same, as does the overall ellipticity distribution across the sample). 100 of these noise realisations were made and we use them to estimate the noise contribution in our measurement, as described in more detail in section \ref{sec:LN_theory:WL}.

\begin{figure}
		\centering
        \includegraphics[scale=0.3]
{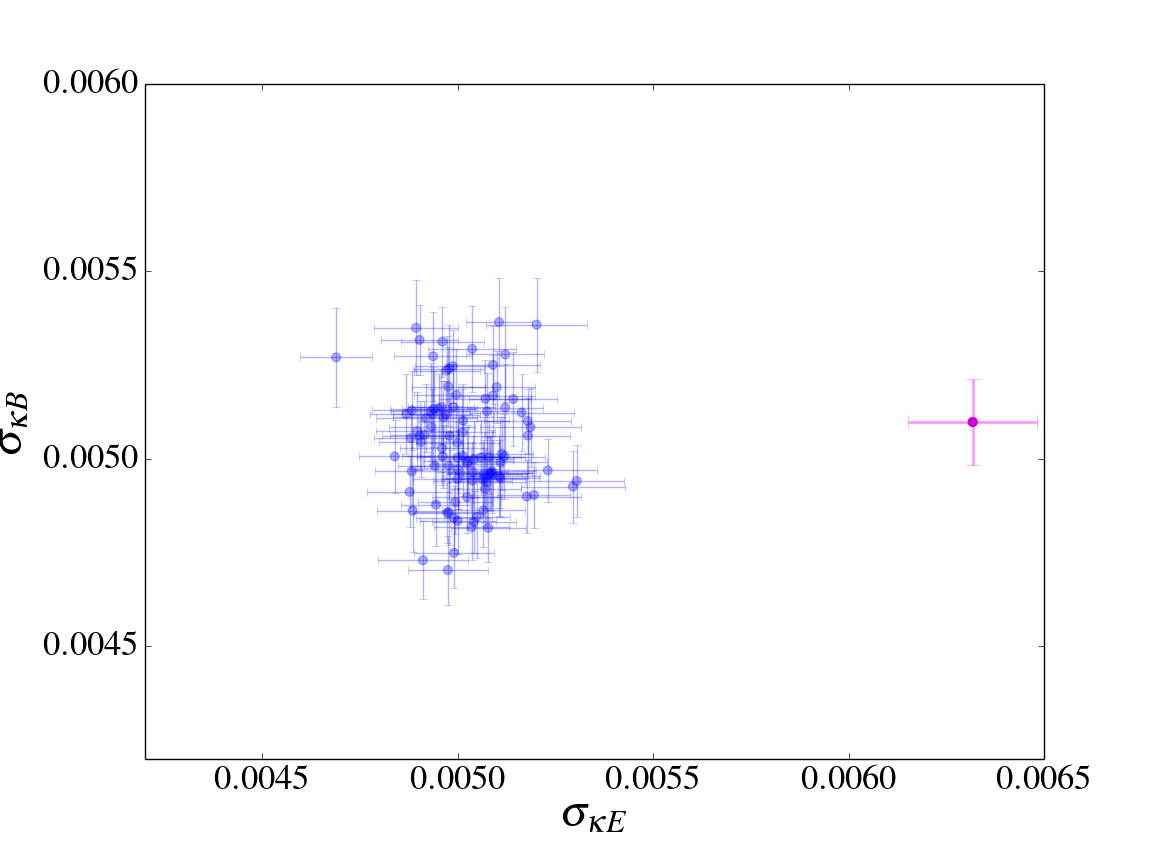}
		\caption{\label{fig:Bmode_SN} Standard deviations of the $\kappa_E$, $\kappa_B$ signal (magenta) and 100 realisations in which the shears have been randomised (blue) at a cell size of 20\arcmin. The random realisations of $\kappa_E$ give an estimate of the shape noise contribution to $\kappa_E$; this accounts for 80\% of the $\kappa_E$ signal. The $\kappa_B$ signal is also a good estimate of the shape noise, with the standard deviation of the $\kappa_B$ signal agreeing with the rms standard deviation of $\kappa_E$ random realisations within 2\%. These standard deviations are calculated via CiC and errors are from jackknife sampling.}
\end{figure}

Fig. \ref{fig:Bmode_SN} shows the standard deviations of the $\kappa_E$, $\kappa_B$ signal (magenta) and 100 noise realisations (blue) for a cell size of 20\arcmin. This shows that the shape noise (given by the random realisations of $\kappa_E$) accounts for 80\% of the $\kappa_E$ signal, underlining the importance of accounting for shape noise in our modelling (as described in section \ref{data:DES_K}). The shape noise dominates the signal most at small scales, accounting for 89\% of the signal at 10\arcmin\ and dropping to 64\% at 40\arcmin. We can see that the $\kappa_B$ signal is also a good estimate of the shape noise, with the standard deviation of the $\kappa_B$ signal agreeing with the rms standard deviation of $\kappa_E$ random realisations within 2\%. 
%This discrepancy in the two estimates of shape noise increases with scale, agreeing within 5\% at 40\arcmin. 
These standard deviations are calculated via CiC (see section \ref{sec:consistency_second_moments} for a prescription for calculating moments from CiC) and errors are from jackknife sampling (see section \ref{sec:method}).

\subsection{MICE Simulations}
\label{sec:data_MICE}

We validate our measurement of CiC from DES SV data using a special set of mock catalogues produced from N-body simulations for the DES collaboration. These come from the Marenostrum Institut de Ci\`{e}ncies de l'Espai Grand Challenges (MICE-GC hereafter) lightcone N-body simulation and associated halo catalogue.

These simulations have been used to produce mock galaxy catalogues for $\sim$200 million galaxies over 5000 deg$^2$ up to a redshift of $z=1.4$. There are also shear estimates for each galaxy made by ray-tracing through the N-body simulations. Every galaxy has a $\kappa_{\rm{WL}}$ value assigned from the integrated dark matter field. 

The simulations are made with $4096^3$ particles of mass $2927 M_{\odot} h^{-1}$ in a box of side 3072 $h^{-1}$Mpc. The MICE-GC has an assumed flat $\Lambda$CDM cosmology: $\Omega_{\rm m} = 0.25$, $\Omega_{\rm b} = 0.044$, $\Omega_\Lambda = 0.75$, $\sigma_8 = 0.8$, $h=0.7$, $n_s = 0.95$. The MICE-GC DES mocks approximately reproduce the magnitude limits of the DES survey and are complete down to apparent magnitude $i < 22.0$ at $z=0.5$.% and $i<24$ at $z=1.0$. 

For use in this paper we have reduced the effective number densities in the mock galaxy and shear catalogues to reflect the statistics of the DES SV samples as well as normalising the redshift to reflect the distribution shown in Fig. \ref{fig:z_kernels}. Each mock catalogue is projected onto a \texttt{HEALPix} map of $N_{side}=8192$, which is then degraded to match the resolution of our data maps where appropriate. 

% It should be noted that the intention in this work is not to quantitatively compare statistics from the MICE simulations and DES data. The reason for carrying out the analysis on MICE is to validate our methods and confirm that we can recover quantities of interest. Differences in the samples and different noise properties will mean that statistics such as the variance are not expected to match those of the DES data.

\begin{figure}
		\centering
        \includegraphics[scale=0.4]{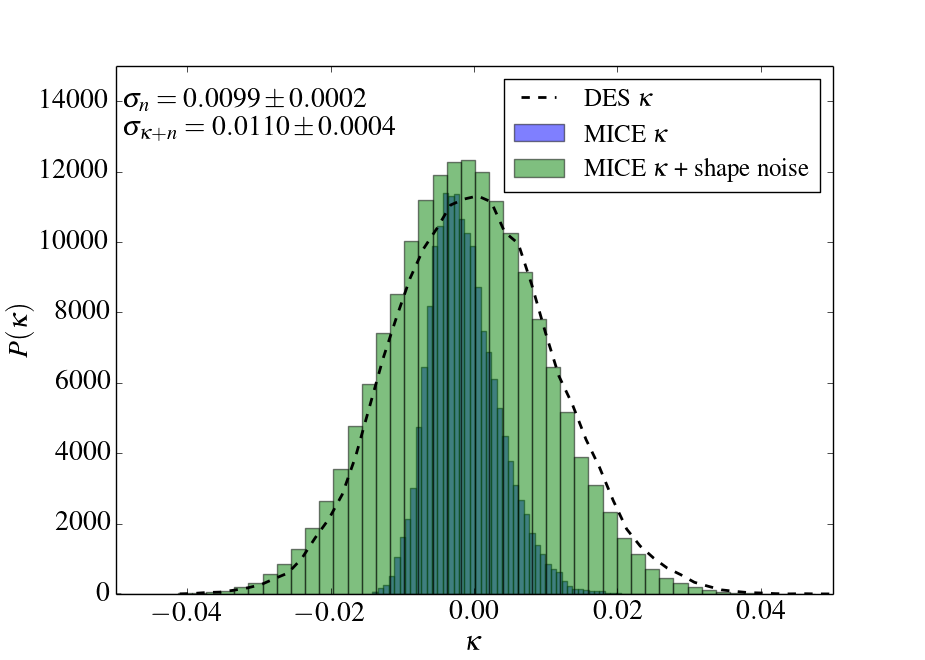}
		\caption{\label{fig:shape_noise_MICE} Distribution of MICE $\kappa_{\rm{WL}}$ at an angular scale of 10\arcmin when DES-like shape noise is added. An estimate of the width of the shape noise distribution is obtained by fitting a Gaussian to the 100 random realisations of DES $\kappa_{\rm{WL}}$. A noise contribution drawn from a Gaussian of this width is added to MICE $\kappa_{\rm{WL}}$ at the level of the cells used to construct the CiC distribution. The darker, narrow histogram is that of the shape noise free MICE $\kappa_{\rm{WL}}$; the lighter histogram shows the distribution once the Gaussian shape noise is added; the black dashed line shows the observed distribution of DES $\kappa_{\rm{WL}}$. The Gaussian width of the DES shape noise estimate, $\sigma_n$ is 0.0099 at this scale, which is 89\% of the width of the resulting noisy MICE $\kappa_{\rm{WL}}$ distribution. }
\end{figure}

In order to be able to compare the distribution of DES $\kappa_{\rm{WL}}$ (which we know has a significant shape noise contribution) with simualtions, we create a second MICE $\kappa_{\rm{WL}}$ sample that has DES-like shape noise added. An estimate of the width of the shape noise distribution is obtained by fitting a Gaussian to the 100 random realisations of DES $\kappa_{\rm{WL}}$. A noise contribution drawn from a Gaussian of this width is added to MICE $\kappa_{\rm{WL}}$ at the level of the cells used to construct the CiC distribution. 

Figure \ref{fig:shape_noise_MICE} shows the effect of adding shape noise to MICE $\kappa_{\rm{WL}}$ in this way at an angular scale of 10\arcmin. The darker, narrow histogram is that of the shape noise free MICE $\kappa_{\rm{WL}}$; the lighter histogram shows the distribution once the Gaussian shape noise is added; and the black dashed line shows the distribution of DES $\kappa_{\rm{WL}}$. At a smoothing scale of 10 arcmin the Gaussian width of the DES shape noise estimate is 0.0099, which is 89\% of the width of the resulting noisy MICE $\kappa_{\rm{WL}}$ distribution; this falls to 63\% at 40\arcmin.

\section{Method}
\label{sec:method}

\subsection{Constructing PDFs via Counts-in-Cells}

% \subsection{}
% \label{sec:cic_method}

The CiC approach is a relatively simple way to measure the distribution of galaxies in a survey, but it is a surprisingly powerful tool. A general CiC distribution for galaxies can be denoted by $f(N,V)$, the probability of finding $N$ galaxies in a volume of space $V$. This can be a 3D volume or, as is the case in this paper, a 2D area on the sky where we count over a population projected along the line of sight. Repeating this procedure with cells of varying radii, $r$, gives us the distribution $f_{r}(N)$, where the moments of $f_{r}(N)$ are related to the volume integrals of the correlation functions of our underlying observable \citep{Peebles1980,Fry1985,Saslaw2000,Fry1994}.

We perform our CiC analysis on the galaxy density contrast and weak lensing covergence maps with \texttt{HEALPix} pixelisation of resolution $N_{\text{side}}=4096$, which corresponds to an average pixel size of 0.9\arcmin. For galaxies, to construct the PDF we sum the galaxy counts, $N$, inside 2D circular cells of fixed radius $r$ in the range 10--40\arcmin. At the median redshift, $z=0.3$, of the sources considered this corresponds to physical scales of 3--10 Mpc. The smallest cells used are 10 times larger than the \texttt{HEALPix} pixels in order to minimise edge effects, and this also avoids any difference in counts across our survey area due to the changing geometry of the \texttt{HEALPix} pixels (see Appendix \ref{sec:Ap_sampling} for a discussion of this assumption). We chose to use randomly positioned circular cells rather than using the \texttt{HEALPix} pixels themselves as this allows us to repeat the analysis straightforwardly at any smoothing scale, rather than using only the fixed scales of \texttt{HEALPix} pixels. The criterion for accepting a cell is that 80\% of its area should fall in unmasked regions (again see Appendix \ref{sec:Ap_sampling} for discussion of this choice). We want to use enough cells that all pixels in the map are covered at least once, and find that this is achieved when the total area of the cells is 20 times that of the survey. We use a coverage of 100 times the total area.

Histograms of the counts give us the distribution $f(N)$, and this procedure is repeated with cells of different radii to obtain the distribution $f_{r}(N)$. Double counting of pixels is accounted for by jackknife errors on the height of each bin in the resulting histogram of counts.  We divide the survey area into 152 approximately equal area (1 deg$^2$) jackknife patches. For a fixed set of randomly generated cells, and removing one patch at a time, we re-make the galaxy and convergence PDFs and re-calculate the statistics of interest in order to produce covariances. 

% The normalised covariance matrix of bin heights of DES $\kappa$ at a smoothing scale of 10 arcmin is shown in figure 
% It can be seen that the correlation is significant only for points separated by less than 5 bins. If we re-bin taking the average over five bins to fit the lognorml model, we can then assume that errors are uncorrelated, using jackknife errors $\sigma = \sqrt{\text{diag(C)}}$.

We repeat our CiC analysis on the DES reconstructed $\kappa$ maps. The `count' in each cell is now the average of the weak lensing convergence $\kappa$ in pixels contained in that cell. 

In Appendix \ref{sec:Ap_syst} we test the impact of spatially varying systematic effects on the DES $\delta_g$ and $\kappa_{\rm{WL}}$ CiC distributions.

It is straightforward to generalise our CiC method to more than one observable. We simply throw the same circles onto each map (using the same mask for each), allowing us to compare counts at the same position for different observables. 

\subsection{Fitting the PDFs}

We fit the lognormal models described in section \ref{sec:LN_theory} to these distributions. For the MICE and DES galaxy density contrast distributions we fit a Poisson sampled lognormal using equation \ref{eq:PxLN}. For MICE $\kappa_{\rm{WL}}$, which has no shape noise, we fit a plain lognormal model (equation \ref{eq:WL_LN}). For the $\kappa_{\rm{WL}}$ distributions which include shape noise (i.e. DES $\kappa_{\rm{WL}}$ and the MICE $\kappa_{\rm{WL}}$ to which we add shape noise), we use equation \ref{eq:GxLN}. 

\begin{figure}
		\centering
        \includegraphics[scale=0.33]{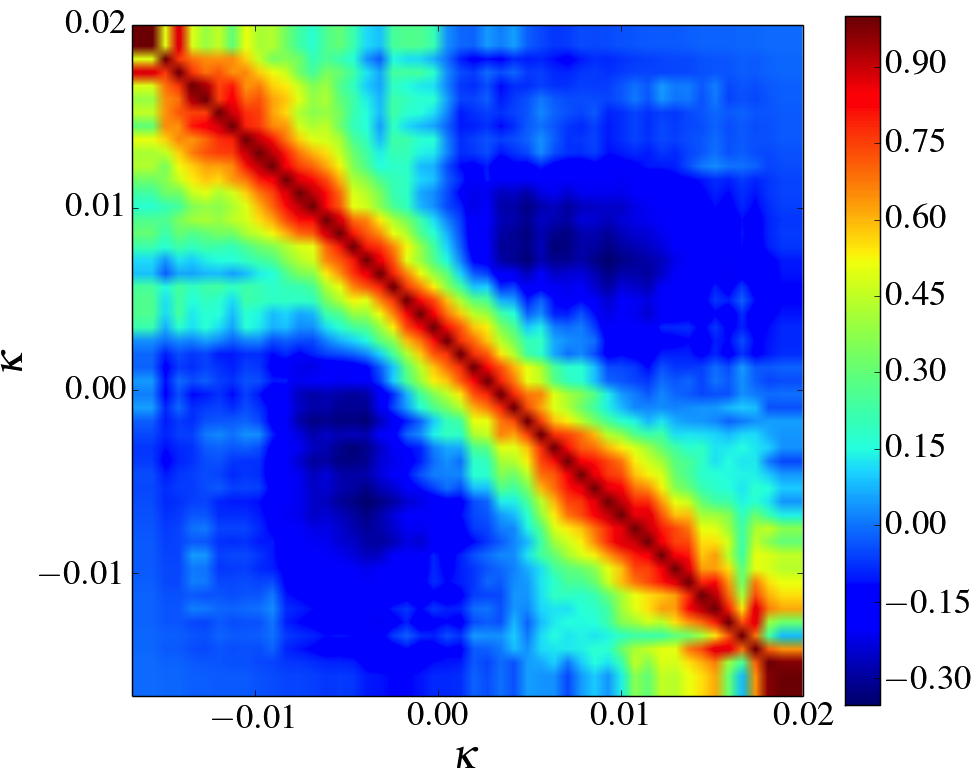}
		\caption{\label{fig:covmat} Correlation matrix of bin heights for a histogram of DES $\kappa_{\rm{WL}}$, at a smoothing scale of 10\arcmin. Derived from jackknife sampling of the DES $\kappa_{\rm{WL}}$ map. }
\end{figure}

The histogram bins in $\delta_g$ or $\kappa_{\rm{WL}}$ are correlated. This is demonstrated in figure \ref{fig:covmat}, which shows the correlation matrix of bin heights of DES $\kappa_{\rm{WL}}$ at a smoothing scale of 10\arcmin. In fitting the lognormal model we take into account these correlations by minimising

\begin{equation}
\chi^2 = (\vec{f} - \vec{d})\mathcal{C}^{-1}(\vec{f} -\vec{d})'.
\end{equation}
Here $f$ is the data vector of the lognormal fit at the bin centres, $d$ is the data vector of bin heights, and $\mathcal{C}$ is the covariance matrix. We remove weak eigenvectors of the covariance matrix via singular value decomposition. % All goodness of fit figures quoted in this work are reduced $\chi^2$ given by $\chi^2$/dof, where the degrees of freedom are the number of singular values used.

% In this paper we first turn our dataset (DES galaxy overdensity, DES WGL $\kappa$ or SPT $\kappa$) into a high resolution healpix map of Nside = 4096, corresponding to a cell size of $\sim 0.74$arcmin$^2$. 

% We then place circular cells of radius $r$ onto the map. The ``count'' in a given cell is then the sum of the quantity in each cell whose centre is inside our cell. If the cell size was similar to the underlying healpix pixel size we would suffer from severe edge effects as some cells are counted which have large areas outside our cell and vice versa. To avoid these edge effects we impose a lower limit on the cell radius of ??? which ensures that the cell area is always ??? larger than the average pixel size. This also avoids any difference in counts across our survey area due to the changing geometry of the healpix pixels. See Appendix \ref{sec:app_cells} for more details.

% \begin{figure}
% 		\centering
%         \includegraphics[scale=0.4]{hist_MICE_WL_khat.png}
% 		\includegraphics[scale=0.4]{hist_DES_WL_khat.png}
% 		\caption{\label{fig:LNfit_WL_DES_35arcmin} Examples of one dimensional Probability Density Functions (PDFs) of $\hat{\kappa}_{WL}$ for MICE and DES at smoothing scale of 15 arcmin. [The latter will be updated once we have final data products from the mass-mapping subgroup.]}
% \end{figure}

% [Paragraph on equivalent formalism for $kappa_{gal}$ to be added.]

\section{Validating Methods on MICE}
\label{sec:MICE}

In this section we verify the methods used to test the lognormality of DES $\delta_{g}$ and $\kappa_{\rm{WL}}$ fields. After checking that the MICE $\delta_{g}$ field is lognormal as we would expect, we see if this is true of the noise-less convergence field. 

To enable easier comparison with the DES $\kappa_{\rm{WL}}$ results we also look at the distribution of the simulation $\kappa_{\rm{WL}}$ for the MICE sample with number of galaxies and $n(z)$ matched to our DES sample, and with DES-like shape noise added. We then look at the joint distribution of $\delta_g$ and $\kappa_{\rm{WL}}$, for the cases with and without shape noise. 
%To confirm that our method of constructing the PDFs via CiC is correctly capturing statistical information we compare the second moments, as calculated via CiC, with the theoretically predicted values derived from power spectra. We also compare these CiC

As an additional check of the validity of the lognormal model, we compare the second moments of the distributions as calculated via CiC with those derived under the assumption of lognormality.
%as a check of the validity of the lognormal model. Satisfying these checks gives allows us to confidently apply these methods to the DES data.

\subsection{Testing Lognormality of MICE Density and Convergence Fields}
\label{sec:LN_MICE}

\subsubsection{One-dimensional PDFs and log-normal fits}
\label{sec:1D_MICE}

\begin{figure*}
		\centering
\includegraphics[scale=0.4] {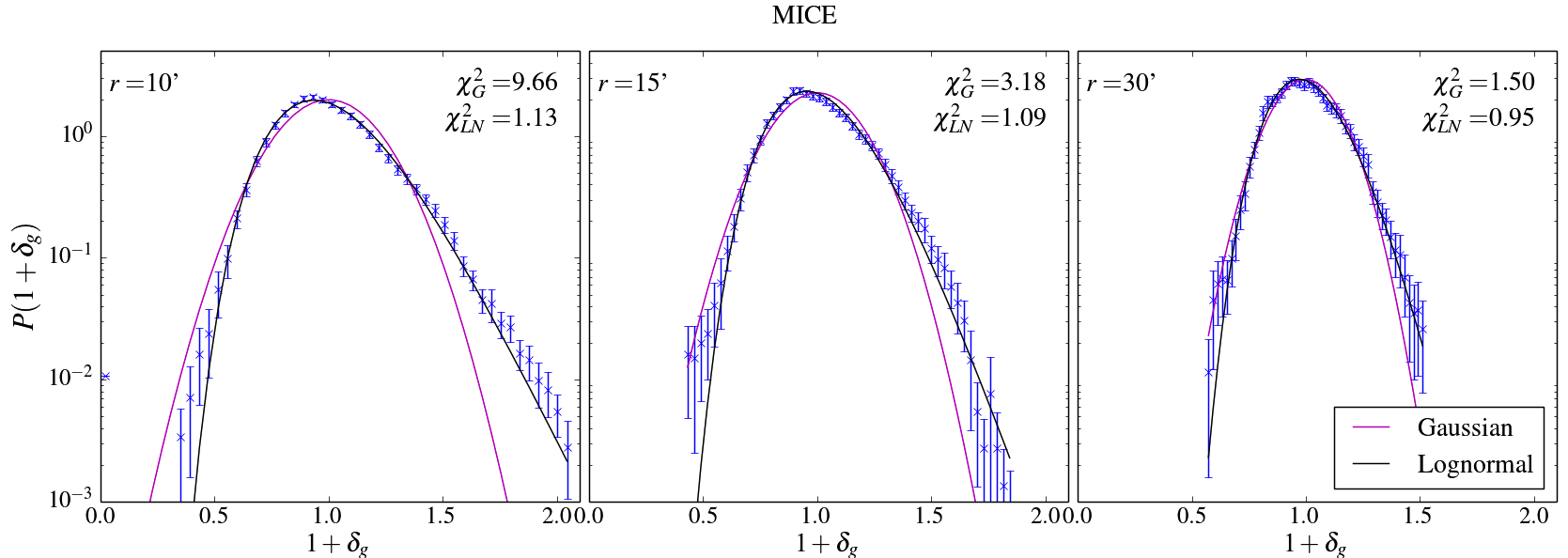}
\includegraphics[scale=0.4]    
% {MICE_kappa_G_LN_cov_r_10_20_30_log_yaxis_svd.png}
{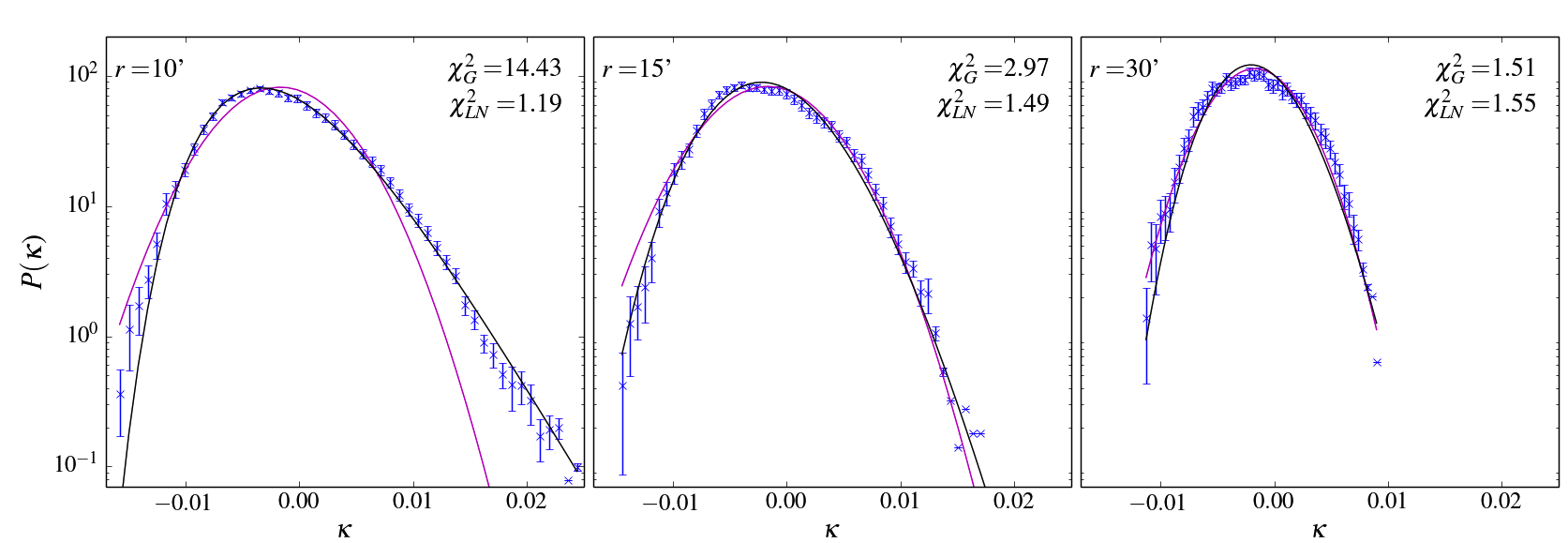}
\includegraphics[scale=0.4]
% {MICE_Knoisy_unmasked_G_conv_cov_r_10_20_30_log_yaxis_svd.png}
{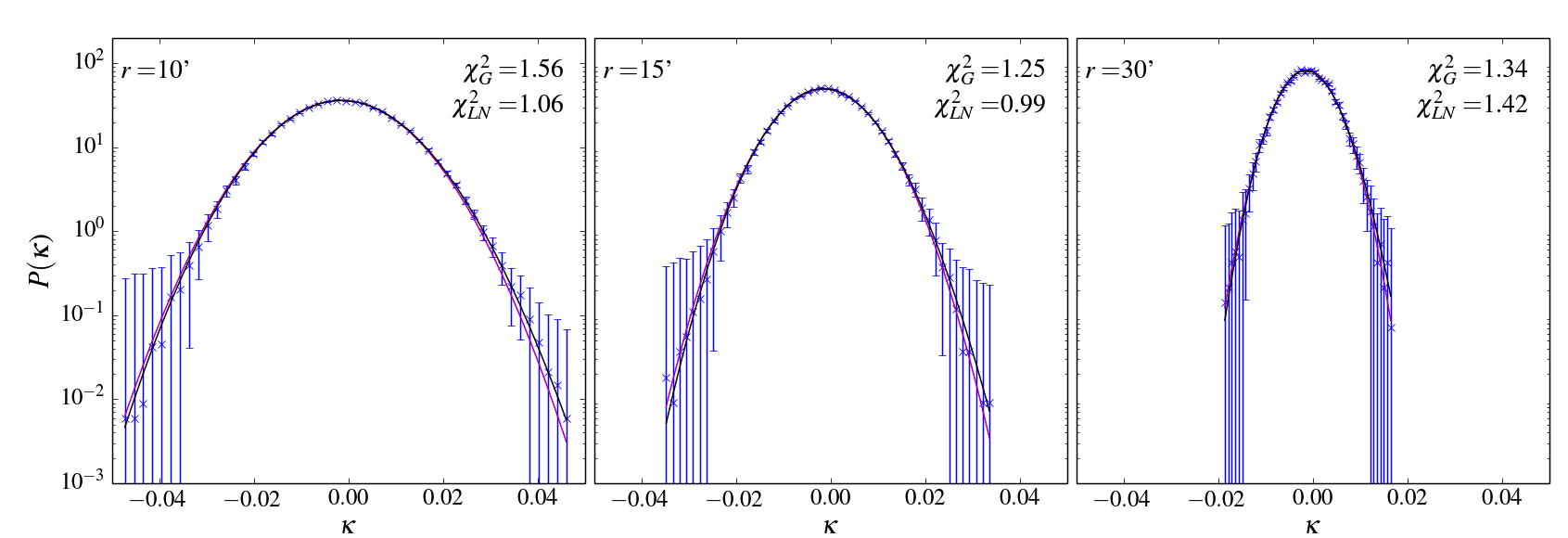}
\caption{\label{fig:gal_K_1Dfit_r5_MICE} \textbf{UPPER ROW:} measured 1D PDF of MICE galaxies at a smoothing scales of 10, 15 and 30\arcmin. The Poisson sampled lognormal fit (black) provides a better fit to the galaxy CiC distribution than the Gaussian (magenta) at a scale of 10\arcmin. The distribution becomes increasingly Gaussian at larger scales. \textbf{MIDDLE ROW:} same as above but for the MICE $\kappa_{\rm{WL}}$ PDF. Again the lognormal provides a good fit at the smallest scale, with the $\kappa_{\rm{WL}}$ distribution becoming more Gaussian at larger scales. \textbf{BOTTOM ROW:} Fits to $\kappa_{\rm{WL}}$ using the sub-sample of MICE with DES-like galaxy density and $n(z)$, and to which DES-like shape noise has been added. This shape noise makes the distribution of $\kappa_{\rm{WL}}$ more Gaussian at all scales. All $\chi^2$ are per degree of freedom.}
\end{figure*}

We first construct a simple histogram of $\delta_{g}$ from the CiC to estimate the 1D PDF of $\delta_{g}$. The histogram uses 50 bins and we calculate jackknife errors on the bin heights as described in the previous section. The result for cells of radius of 10, 15 and 30\arcmin\ is shown in the upper panel of Fig. \ref{fig:gal_K_1Dfit_r5_MICE}. %The lower panel shows the equivalent 1D PDF of the convergence field. %For reference, a cell size of %[insert examples of angular $-->$ physical distances]. 
We fit a Poisson sampled lognormal distribution as described in eqn. \ref{eqn:LN} with $w$ as the single free parameter. The best-fit lognormal, which minimises $\chi^2$, is shown as a solid black line and the best fit Gaussian (magenta) is shown for comparison. At a cell size of 10\arcmin\ (corresponding to about 3 Mpc at the median redshift $z=0.3$) it is clear that the lognormal model fits the data better, reflecting the non-linear clustering at this scale. The counting of information inside a cell can be thought of as a form of smoothing where the cells form a top-hat filter with a fixed size. As the size of our cells increases we average information on increasingly large scales and lose sensitivity to the effects of non-linear clustering on small scales. 

The lognormal distribution is designed to capture some of the information present as a result of non-linear evolution, so we would expect it to become less pronounced as the effective smoothing scale increases. This is indeed the case: at a cell radius of 10\arcmin\ the lognormal model is highly favoured, with a $\chi^{2}/DOF = 1.13$, compared to 9.66 for the Gaussian. At a cell size 30\arcmin\ (corresponding to a physical scale of 8Mpc at the median redshift) the distribution has become much more Gaussian with best-fit $\chi^{2}$/DOF for the Gaussian model now 1.50. The lognormal model is still favoured at this scale, with best-fit $\chi^{2}$/DOF $=0.95$. 

The result for the MICE $\kappa_{\rm{WL}}$ PDF is shown in the middle panel of fig. \ref{fig:gal_K_1Dfit_r5_MICE}. Since there is no shape noise in the simulation we fit a plain lognormal, shown by the black line. As discussed in section \ref{sec:LN_theory:WL}, in order to fit a lognormal model to $\kappa_{\rm{WL}}$ one must assign a value to $\kappa_0$, the minimum convergence parameter in equation \ref{eq:WL_LN}. At 10\arcmin\ we jointly fit $\kappa_{0}$ and the lognormal width in equation \ref{eq:WL_LN}, finding best-fit $\kappa_{0}=0.021$. For larger scales we find that it is not possible to jointly constrain $\kappa_0$ and the width of the lognormal as they are degenerate. We therefore use the theoretically derived $\kappa_0=0.050$, described in section \ref{sec:LN_theory:WL}. 

Since the convergence is the weighted sum of the mass fluctuations along the line of sight we expect it to be only approximately lognormal. At a smoothing scale of 10\arcmin\ the lognormal is a good fit, with $\chi^2$/DOF$ = 1.19$, and it is significantly preferred to the Gaussian model, which has a best-fit $\chi^2$/DOF$ = 14.43$. This lognormality of $\kappa_{\rm{WL}}$ at small scales is in line with \cite{Taruya2002} who found that a lognormal model was a good fit to simulated $\kappa_{\rm{WL}}$ at angular scales of 2 - 4\arcmin. Increasing the cell radius above 10\arcmin\ removes the clear preference for the lognormal, and the lognormal and Gaussian models fit the data equally well at cell radii of 30\arcmin. The fixed, theoretically derived $\kappa_0 = 0.050$ allows the lognormal model with a single free parameter to fit the distribution well at 15\arcmin, but at larger cell radii this model does very slightly worse than the Guassian model. This suggests that this value of $k_0$ may not be a good estimate for the minimum $\kappa$ in the CiC PDF for larger cells. This makes sense as this $\kappa_0$ corresponds a pure void along the line of sight, which is a decreasingly likely observation as the cell radius increases.

The final row of figure \ref{fig:gal_K_1Dfit_r5_MICE} shows the distribution of $\kappa$ using the sub-sample of MICE with DES-like galaxy density and $n(z)$, and to which DES-like shape noise has been added, as described in section \ref{sec:data_MICE}. The shape noise dominates the resulting distribution, particularly at smaller scales. The width of the distribution of shape noise is 74\% of the width of the noisy $\kappa$ distribution at 40\arcmin, and at 10\arcmin\ it accounts for 89\%.  We model the noisy $\kappa$ distribution with a lognormal convolved with Gaussian noise as described in section \ref{sec:LN_theory:WL}, using equation \ref{eq:GxLN}. Again we find that it is not possible to jointly constrain $\kappa_0$ and the width of the lognormal at scales above 10\arcmin\ as they are degenerate. We therefore use the theoretically derived $\kappa_0=0.049$. It can be seen from figure \ref{fig:gal_K_1Dfit_r5_MICE} that at all scales the shape noise makes the noisy $\kappa$ distribution much more Gaussian. 
%This increased Gaussianity is most prominent at smaller scales where the shape noise dominates the signal more than it does at larger scales. The lognormal convolved with Gaussian noise provide a good fit at all scales with best-fit $\chi^2$ = 0.87, 1.00, 0.90 at scales of 10, 20 and 30 arcmin respectively. These are broadly in line with or lower than the  $\chi^2$ for the Gaussian model, which are 1.32, 1.01, 1.67 at 10, 20 and 30 arcmin. 

Despite the low signal to noise, at 10\arcmin\ the lognormal convolved with Gaussian noise provides a better fit than the simple Gaussian, with $\chi^2$/DOF$ = 1.06$ and 1.56 respectively. At scales larger than this the Gaussian model performs as well as the lognormal. As with the noise free convergence distribution, the theoretically derived $\kappa_0$ seems to be a less suitable choice at larger scales as the Gaussian model provides a better fit for scales above 30\arcmin.

\subsubsection{Joint galaxy-convergence distribution}
\label{sec:Joint_MICE}

In this sub-section we study the joint distribution of galaxy overdensities and weak lensing convergence and determine to what extent it can be described as a bivariate lognormal distribution. We look at joint distributions using both the full MICE sample, and the subsample with DES-like galaxy density and $n(z)$ and the addition of DES-like shape noise. As in the 1D case, the full sample with higher galaxy density allows us to better capture any lognormal behaviour, and the DES-like sample allows us to compare the results for DES data given in the next section with simulations.

We can make a simple quantitative estimate of the relative correlation of $\delta_{g}$ and $\kappa_{\rm{WL}}$  by calculating the Pearson product-moment correlation coefficient, $r$, for the joint PDF, where
\begin{equation}
\rho_{X,Y} = \frac{\textrm{Cov}(X,Y)}{\sigma_{X}\sigma_{Y}} = \frac{\left< (X-\bar{X}) (Y-\bar{Y})\right>}{\sigma_{X}\sigma_{Y}} .
\end{equation}

\begin{figure}
		\centering
		\includegraphics[scale=0.37]{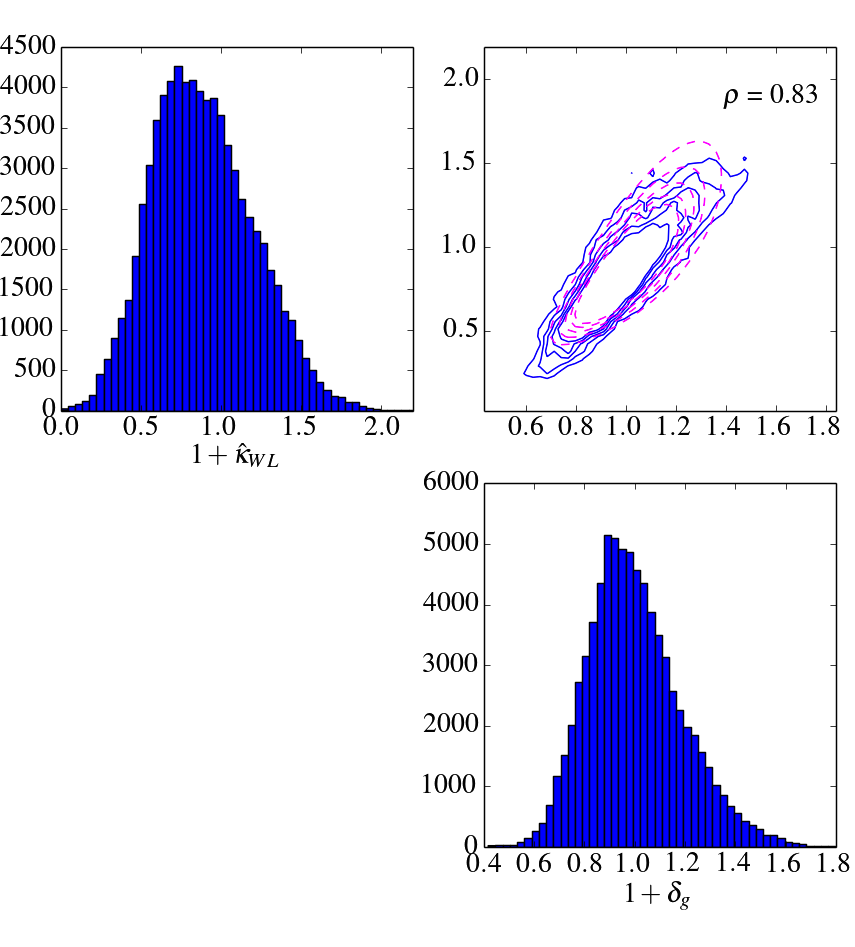}
        		\includegraphics[scale=0.37]{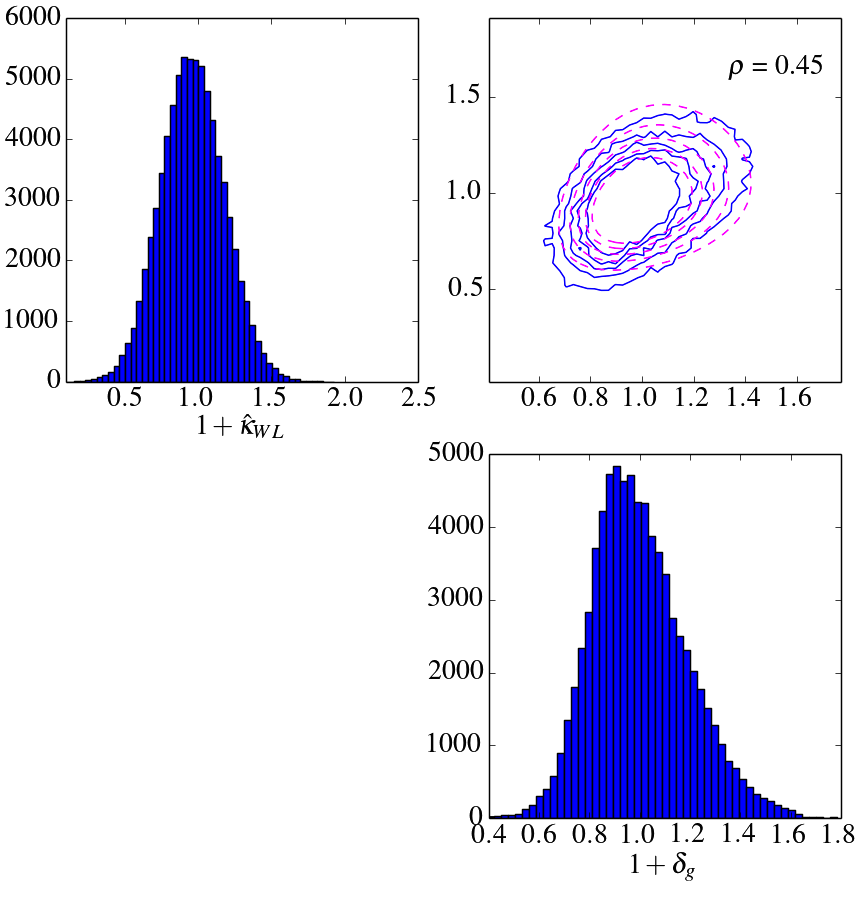}
		\caption{\textbf{UPPER PANEL:} Joint CiC distribution of weak lensing convergence and galaxy density contrast for MICE simulation at a smoothing scale of 15\arcmin. The top right plot shows the bivariate lognormal fit to MICE simulations. Contours for the simulation are given by the solid blue lines, with dashed magenta contours for the fit. Also shown are the 1D PDFs for $1+\delta_g$ and $1+\hat{\kappa}_{WL}$ individually. PDFs are calculated via the CiC method with cells of radius 15\arcmin. As in the rest of this paper, galaxies are selected over the redshift range $0.1 < z < 0.5$ and WL sources are restricted to the range $0.6 < z < 1.3$. This joint distribution has a Pearson correlation coefficient of $r = 0.83$. \textbf{LOWER PANEL:} Same but with DES-like shape noise added to $\kappa_{\rm{WL}}$. The Pearson correlation coefficient drops to 0.45 with the addition of this shape noise. }
\label{fig:2Dscatter_MICE}
\end{figure}

% \begin{figure}
% 		\centering 
% 		\includegraphics[scale=0.4]{MICE_2Dfit_g_0p2z0p5_K_0p6z1p3_r15_cond_prob.png}
% 		\caption{Fit of bivariate lognormal fit to MICE simulation data at a smoothing scale of 15 arcmin. Contours for the data are given by the solid blue lines, with dashed magenta contours for the fit. Best fit values of the three free parameters are given. Note that $r$ here is not equivalent to the Pearson correlation coefficient given in the previous plot.}
%         \label{fig:2D_LN_fit}
% \end{figure}

% We do not expect full correlation of $\kappa_{WL}$ and $\delta_g$ because there is a different window function on the galaxy projection versus that on $\kappa_{WL}$.

We begin with the joint distribution of $\delta_{g}$ and $\kappa_{WL}$ with no shape noise, which is shown in the upper panel of Fig. \ref{fig:2Dscatter_MICE} for a smoothing scale of 15\arcmin. % This is just a case of taking each cell in turn and plotting the $\delta_{g}$ value against the $\kappa_{WL}$ value. Projecting these points onto either axis recovers the relevant 1D PDF.
The blue contours in the top right section of this plot show the joint PDF, and the dashed magenta contours show the bivariate lognormal fit. 
%We fit our joint PDF with the bivariate lognormal, $f(\delta,\kappa)$, finding the best-fit values for three free parameters: $\{ \omega_{\delta}, \omega_{\kappa}, \omega_{\delta\kappa} \}$. 
Since there is no shape noise in this case the bivariate fit is given by equation \ref{eq:biv} but omitting the Gaussian convolution. 
%We show an example best-fit 2D lognormal in Fig. \ref{fig:2Dscatter_MICE} 
We expect the correlation coefficient $\rho$ to be high (close to one) since the galaxies considered are responsible for the lensing. This is indeed what we see: the Pearson correlation coefficient is 0.81 at a smoothing scale 10\arcmin\ and 0.89 at 40\arcmin. We do not see full correlation because the relevant window functions - the lensing efficiency function of the source sample and the galaxy redshift distribution of the galaxy sample - do not overlap precisely.

The lower panel of this figure shows the case where MICE $\kappa_{\rm{WL}}$ has had shape noise added. This noise reduces the correlation of the $\kappa_{\rm{WL}}$ with $\delta_g$, smearing out the joint distribution (shown on the top right of the figure) versus the shape noise free case. The Pearson correlation coefficient is reduced to 0.45.

\subsection{Comparison of Moments}
\label{sec:Joint_MICE}

% \begin{figure}
% 		\centering
% 		\includegraphics[scale=0.35]
% % {MICE_theory_data_gg_KK_2panels.png}
% {MICE_mom2_theory_CiC_LN.png}
% \caption{\textbf{UPPER PANEL:} Comparison of the second moments of MICE galaxy density contrast measured directly via CiC (blue), from lognormal fits to the CiC PDF (red) and analytically derived two point correlation functions (black dashed line).
%  \textbf{LOWER PANEL:} same but for MICE weak lensing convergence.}
%  \end{figure}
 
 \begin{figure}
\includegraphics[scale=0.35]
{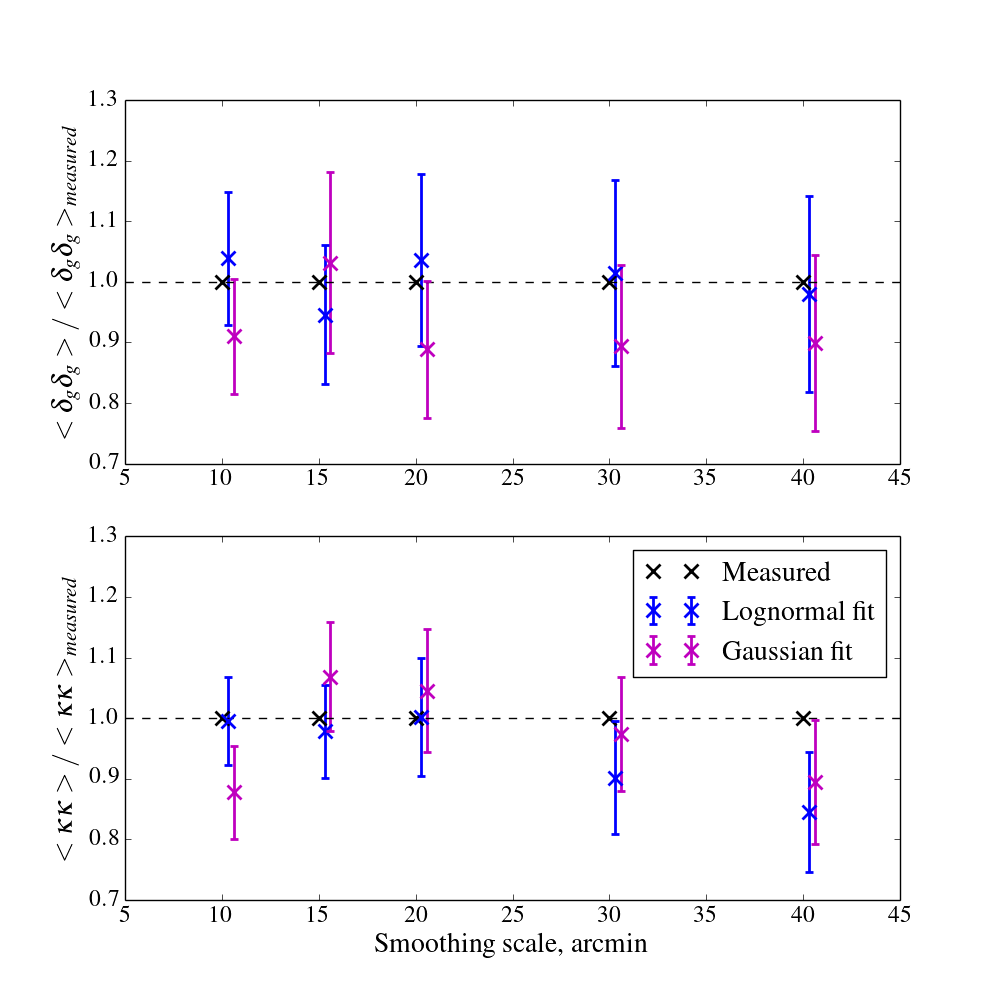}
\caption{\textbf{UPPER PANEL:} comparison of second moments of MICE galaxy density contrast as a function of smoothing scale, directly measured via CiC and from lognormal and Gaussian fits to the CiC PDF. Blue data points show the ratio of the variance $\left< \delta_g^{2} \right>$ from our fits to the 1D lognormal distribution to that calculated directly from the CiC PDF; black data points show the same but for the Gaussian fit. Data points are offset slightly in scale for clarity. \textbf{LOWER PANEL:} same but for shape noise free MICE weak lensing convergence.}
        \label{fig:mom2MICE}
\end{figure}

We can use the second moments to check the validity of the lognormal modelling by comparing the moments derived directly from the CiC with those derived by fitting a lognormal model to the CiC PDF. The second moments of the MICE galaxy and convergence fields $\langle\delta_g^2\rangle$ and $\langle\kappa^2\rangle$ can be calculated from the CiC (as described in Appendix \ref{sec:Ap_mom}). The moments derived under lognormal modelling are given by equations \ref{eq:gg} and \ref{eq:kk}.

First we calculate the variance of the MICE galaxy PDF, shown in the first panel of Fig. \ref{fig:mom2MICE}. Blue data points show the ratio of the variance $\left< \delta_g^{2} \right>$ from fitting a lognormal to the CiC PDF and that calculated directly from the CiC. Errors on $\left< \delta_g^{2} \right>$ directly from CiC are produced by jackknife sampling; errors on the $\left< \delta_g^{2} \right>$ derived from the lognormal fit are from the 1$\sigma$ width of the likelihood of the lognormal width. 

The lognormal model gives a good estimate of the variance of the MICE galaxy density contrast distribution (with Poisson shot noise accounted for) at all scales. It gives a better estimate of the variance than a Gaussian model at all scales, and particularly at 10\arcmin. 
%The only excption is at 15\arcmin where the Gaussian model gives a better estimate, but this is because the Gaussian PDF overestimates the CiC distribution at low $\delta_g$ and underestimate it at high $\delta_g$, resulting in a PDF that happens to have a width very similar to the data, but a poor fit (see middle panel, top row of fig. \ref{fig:gal_K_1Dfit_r5_MICE}). 
The lognormal model also gives a good estimate of the variance of the weak lensing convergence distribution at scales up to 20\arcmin. The poorer estimates at 30 and 40\arcmin\ are due to the fact that we fix $\kappa_{min}$ to the theory value at these scales.

These results suggest that within the ranges of scales discussed, the lognormal model can be used to estimate the two point statistics of both the galaxy density contrast and weak lensing convergence distributions to reasonable accuracy in these simulations.

% First we calculate the variance of the MICE galaxy PDF, shown in the first panel of Fig. \ref{fig:mom2MICE}. Blue data points show ratio of the variance $\left< \delta^{2} \right>$ from our fits to the 1D lognormal distribution to that calculated directly from the CiC PDF, as described above. Error bars are produced by jackknife sampling. The univariate lognormal model with Poisson shot noise gives a good estimate of the variance of the MICE galaxy overdensity at all scales from 5,  beginning to diverge from the value measured directly from the distribution via CiC at a scale of 40 arcmin. The univariate lognormal model provides a good estimate for the variance of the MICE weak lensing convergence distribution for scales below 30 arcmin. This suggests that Within these ranges the lognormal can be used to model the two point statistics of both the galaxy overdensity and weak lensing convergence to reasonable accuracy.

% One way of estimating the effectiveness of the bivariate lognormal description of our joint distribution is to fit a bivariate Gaussian distribution to the same data and compare the goodness of fit. [Figures comparing GoF to be added once we have final data]

\section{Testing lognormality of DES density and convergence fields}
\label{sec:LN_DES}
Here we repeat the analysis of the previous section with DES galaxy and convergence maps, looking first distributions individually and then at their joint distribution.

\subsection{One-dimensional PDFs and log-normal fits}
\label{sec:1D_DES}

% Figure \ref{fig:DES_gal_K_1Dpdf} shows the PDFs of DES galaxy overdensity (upper panel) and convergence (lower panel). The lognormal fit shown in black. Note that smoothing scale is 5 arcmin for the galaxy PDF, and 15 arcmin for the convergence PDF due to the lower resolution of the $\kappa$ map.

% \begin{figure}
% 		\centering
%         \includegraphics[scale=0.27]{DES_gals_G_PxLNfits_r10_g_0p2z0p5_K_0p6z1p2_4096_cov20.png}
%         \includegraphics[scale=0.40]{DES_WL_G_LNfits_r_20_g_0p2z0p5_K_0p6z1p2_4096_cov20Kgal.png}
%         \includegraphics[scale=0.40]{DES_WL_G_LNfits_r_20_g_0p2z0p5_K_0p6z1p2_4096_cov20KWL.png}

% 		\caption{\label{fig:DES_gal_K_1Dpdf} Upper panel: measured 1D PDF of DES galaxies at a smoothing scale of 10 arcmin. The Poisson sampled log-normal fit (black) provides a better fit than the Gaussian (magenta), demonstrating the log-normality of the galaxy CiC distribution at this scale. Middle and lower panels: lognormal fits with Gaussian noise for DES $\kappa$, and $\kappa_{gal}$, with smoothing scale of 20 arcmin. }
% \end{figure}

\begin{figure*}
		\centering
        \includegraphics[scale=0.38]{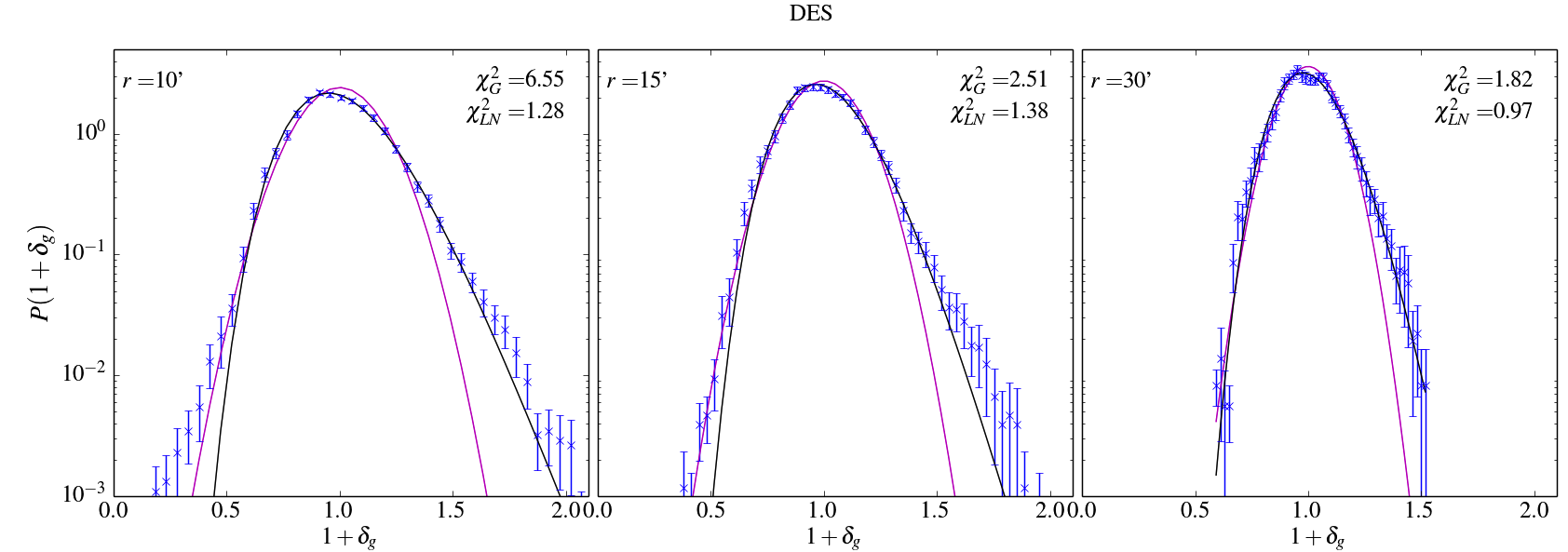}
         \includegraphics[scale=0.38]
% {DES_kappa_G_LNconv_r_10_20_30_log_yaxis_svd.png}
{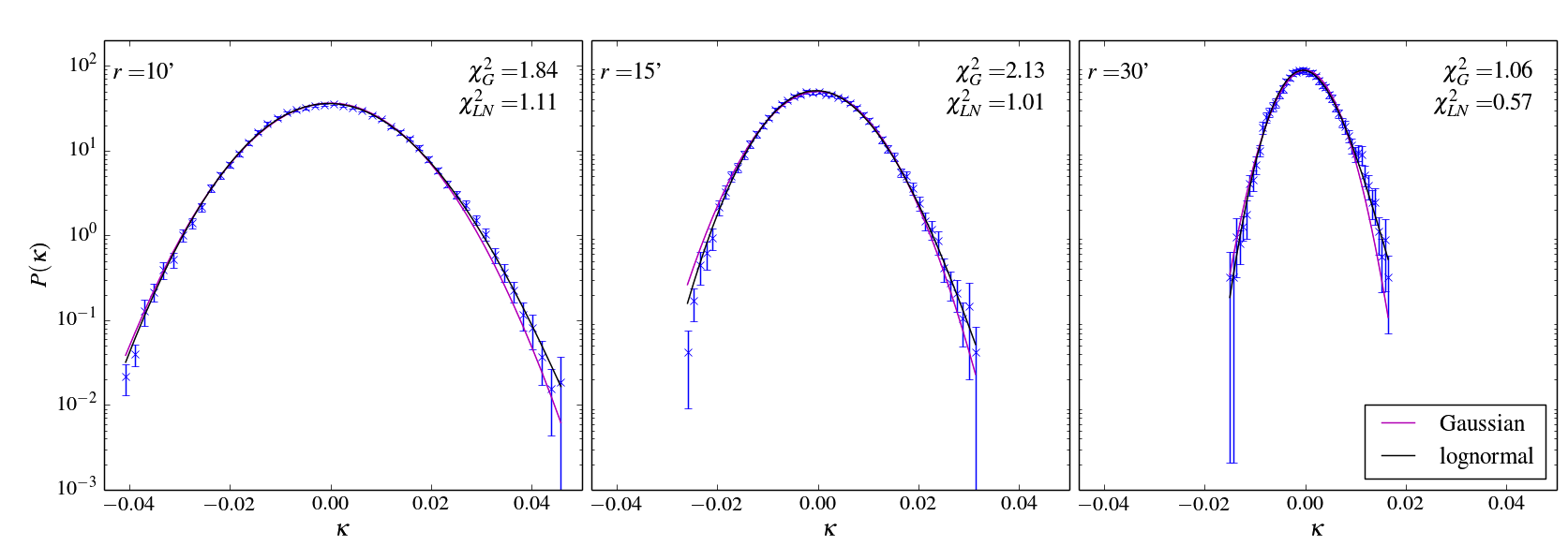}
		\caption{\label{fig:smoothing} \textbf{UPPER ROW}: measured 1D PDF of DES galaxies at a smoothing scales of 10, 15 and 30\arcmin. At 10\arcmin\ the Poisson sampled log-normal fit (black) provides a much better fit than the Gaussian (magenta), demonstrating the log-normality of the galaxy CiC distribution at this scale. At larger scales the distribution becomes more Gaussian. \textbf{LOWER ROW}: same but for $\kappa_{\rm{WL}}$. Here the lognormal model includes Gaussian shape noise, which provides a good fit at all scales. Error bars on the counts PDFs are jackknife errors. All $\chi^2$ are per degree of freedom.}

\end{figure*}

Fig. \ref{fig:smoothing} shows 1D CiC PDFs for the DES galaxy density contrast (top row) and $\kappa_{\rm{WL}}$ (second row) fields for different cell radii. The lognormal fit is again shown in black, and for comparison a Gaussian fit is shown in magenta. 

For the $\delta_g$ PDF at 10\arcmin\ the lognormal model is clearly favoured, with $\chi^2$/DOF$ = 1.28$ compared to 6.55 for the Gaussian model. This confirms the expected lognormal behaviour at non-linear scales, indicating that our CiC procedure is capturing non-linear clustering information beyond the Gaussian assumption at smaller radii. As in the simulations the $\delta_g$ PDFs clearly appear more Gaussian at larger cell radii, although the lognormal model still provides a better fit than the Gaussian even at 30\arcmin, with $\chi^2$/DOF of 0.97 and 1.82 respectively.

% \begin{table}
% \centering
% \begin{tabular}{|c|c|c|c|c|c|}
% \hline
% $r$, arcmin & $\sigma$ & $\chi^2_{G}$ & $\chi^2_{LN}$  & DOF &  $\langle\delta_g\delta_g\rangle$ \\
% \hline
% 10 & 0.184  &  6.55  &  1.28  &11&  $3.44 \pm 0.30 \times 10^{-2}$  \\
% 15 & 0.156  &  2.51  &  1.38  &11&  $2.46 \pm 0.28 \times 10^{-2}$  \\
% 20 & 0.146  &  1.79  &  1.03  &12&  $2.15 \pm 0.29 \times 10^{-2}$  \\
% 30 & 0.126  &  1.82  &  0.97  &17&  $1.60 \pm 0.16 \times 10^{-2}$  \\
% 40 & 0.122  &  1.04  &  0.84  &22&  $1.26 \pm 0.17 \times 10^{-2}$  \\
% \hline
% \end{tabular}
% \caption{Best-fit parameters and derived statistics from lognormal fits to CiC PDFs of DES galaxy overdensity, for varying cell radii. First first column gives the cell radius, and the second column is the width of the best fitting Poisson-sampled lognormal. The following two columns are the $\chi^2$ per degree of freedom for Gaussian and Poisson sampled lognormal fits. The final column is the second moment of the best-fitting lognormal PDF, derived from the lognormal width, with $1\sigma$ errors given by the likelihood of the lognormal width.}
% \label{table:LN_g}
% \end{table} 

\begin{table}
\centering
\begin{tabular}{|c|c|c|c|c|c|}
\hline
$r$, arcmin & $\sigma$ & $\chi^2_{G}$ & $\chi^2_{LN}$  & DOF &  $\langle\delta_g\delta_g\rangle \times 10^{-2}$ \\
\hline
10 & 0.184  &  72.05  &  14.08  &11&  $3.44 \pm 0.30$  \\
15 & 0.156  &  27.61  &  15.18  &11&  $2.46 \pm 0.28$  \\
20 & 0.146  &  21.48  &  12.36  &12&  $2.15 \pm 0.29$  \\
30 & 0.126  &  30.94  &  16.49  &17&  $1.60 \pm 0.16$  \\
40 & 0.122  &  22.88  &  18.48  &22&  $1.26 \pm 0.17$  \\
\hline
\end{tabular}
\caption{Best-fit parameters and derived statistics from lognormal fits to CiC PDFs of DES galaxy density contrast, for varying cell radii. First first column gives the cell radius, and the second column is the width of the best fitting Poisson-sampled lognormal. The following columns are the minimum $\chi^2$ for Gaussian and Poisson sampled lognormal fits, and the number of degrees of freedom. The final column is the second moment of the best-fitting lognormal PDF, derived from the lognormal width, with $1\sigma$ errors given by the likelihood of the lognormal width.}
\label{table:LN_g}
\end{table}

% \begin{table}
% \centering
% \begin{tabular}{|c|c|c|c|c|}
% \hline
% $r$, arcmin & $\sigma$ & $\chi^2_{LN}$ & $\chi^2_{G}$  &  $\langle\delta\delta\rangle$ \\
% \hline
% 10 & 0.184  &  6.55  &  1.28  &  $3.44 $  \\
% 15 & 0.156  &  2.51  &  1.38  &  $2.46 $  \\
% 20 & 0.146  &  1.79  &  1.03  &  $2.15 $  \\
% 30 & 0.126  &  1.82  &  0.97  &  $1.60 $  \\
% 40 & 0.122  &  1.04  &  0.84  &  $1.26 $  \\
% \hline
% \end{tabular}
% \caption{Best-fit parameters and derived statistics from lognormal fits to CiC PDFs of DES galaxy overdensity, for varying cell radii. First first column gives the cell radius, and the second column is the width of the best fitting Poisson-sampled lognormal. The following two columns are the $\chi^2$ per degree of freedom for Gaussian and lognormal fits. The final column is the second moment of the best-fitting lognormal PDF, derived from the lognormal width.}
% \label{}
% \end{table} 

The best-fit values of the free parameters of the lognormal fits to the DES galaxy density contrast distribution, the $\chi^2$, the number of degrees of freedom (DOF) and the second moment of the best-fit lognormal PDF are given in table \ref{table:LN_g}, for smoothing scales of 10 - 40\arcmin. 

% \begin{table}
% \centering
% \begin{tabular}{|c|c|c|c|c|c|}
% \hline
% $r$, arcmin & $\kappa_{0}$ & $\sigma$ & $\chi^2_{G}$ & $\chi^2_{LN}$  &  $\langle\kappa\kappa\rangle$ \\
% \hline
% 10 & 0.021 & 0.235 & 1.84  & 1.11  & $2.44 \pm 0.45 \times 10^{-5}$  \\
% 15 & 0.017 & 0.248 & 2.13  & 1.01  & $1.59 \pm 0.34 \times 10^{-5}$  \\
% 20 & 0.016 & 0.238 & 1.09  & 0.46  & $1.46 \pm 0.29 \times 10^{-5}$  \\
% 30 & 0.009 & 0.314 & 1.06  & 0.57  & $8.41 \pm 2.12 \times 10^{-6}$  \\
% 40 & 0.008 & 0.300 & 1.14  & 0.66  & $7.09 \pm 1.91 \times 10^{-6}$  \\
% \hline
% \end{tabular}
% \caption{Same as Table \ref{table:LN_g} but for DES weak lensing convergence. The lognormal fit accounts for shape noise, so the statistics quoted are for the de-noised $\kappa_{WL}$ distribution. The additional information given vs. Table \ref{table:LN_g} , in the second column, is the best-fit minimum convergence parameter $\kappa_0 = -\kappa_{min}$.}
% \label{table:LN_k}
% \end{table}

\begin{table}
\centering
\begin{tabular}{|c|c|c|c|c|c|c|}
\hline
$r$, arcmin & $\kappa_{0}$ & $\sigma$ & $\chi^2_{G}$ & $\chi^2_{LN}$  & DOF &  $\langle\kappa\kappa\rangle \times 10^{-5}$ \\
\hline
10 & 0.021 & 0.235 & 18.41  & 11.10 &10& $2.44 \pm 0.45 $  \\
15 & 0.017 & 0.248 & 19.17  & 9.09  &9&  $1.59 \pm 0.34 $  \\
20 & 0.016 & 0.238 & 10.92  & 4.63  &10& $1.46 \pm 0.29 $  \\
30 & 0.009 & 0.314 & 11.66  & 6.27  &11& $0.84 \pm 0.21 $  \\
40 & 0.008 & 0.300 & 14.82  & 8.58  &13& $0.71 \pm 0.19 $  \\
\hline
\end{tabular}
\caption{Same as Table \ref{table:LN_g} but for DES weak lensing convergence. The lognormal fit accounts for shape noise, so the statistics quoted are for the de-noised $\kappa_{WL}$ distribution. The additional information given vs. Table \ref{table:LN_g} , in the second column, is the best-fit minimum convergence parameter $\kappa_0 = -\kappa_{min}$.}
\label{table:LN_k}
\end{table}

% $\kappa_{gal}$, shown the middle panel, also becomes more Gaussian with increasing scale, but the lognormal model is a better fit to the data at all scales up to to 40 arcmin. The shift in the lognormal model, $\kappa_0$, was fitted as a free parameter. 

%As discussed in section \ref{sec:LN_theory:WL}, 
%In order to fit a lognormal model to $\kappa$ one must assign a value to
The second row of Fig. \ref{fig:smoothing} shows the DES $\kappa_{\rm{WL}}$ distribution. We find that it is possible to jointly constrain $\kappa_0$, the minimum convergence parameter in equation \ref{eq:WL_LN}, and the width of the lognormal at all smoothing scales. The best-fit values of $\kappa_0$ as well as the lognormal width $\sigma$, best-fit $\chi^2$ and the second moment of the best-fit lognormal PDF are given in table \ref{table:LN_k}. The best-fit $\kappa_0 = 0.021$ and $\sigma = 0.235$ at cell radius 10\arcmin\ are in good agreement with the results from the MICE simulation at this scale, which are 0.023 and 0.226 respectively. Note that for larger scales we fix $\kappa_0$ at the theory value of 0.05 in the simulations, so would not expect close agreement of the best-fit lognormal width with that of the data at the these scales.

% Both Taruya et al. 2002 and Hilbert et al. 2011 extract $k_0$ from simulations but neither probe scales as large as the minimum scale we are able to probe, which is cell radius 10 arcmin or angular scale 20 arcmin. The largest scale at which $\kappa_0$ is measured by Taruya et al. is 10 arcmin, and depending on the assumed cosmology their $\kappa_{min}$ (where $\kappa_0 = -\kappa_{min}$) is -0.015 - 0.030 for source redshift 1. Accounting for the increasing trend in $\kappa_{min}$ shown in Fig. 1 of Taruya et al. and a slight additional expected decrease (Hilbert et al. 2011) due to the lower source redshift of our sample ($z=0.89$), our best-fit $\kappa_{min} = -0.021$ at cell radius 10 arcmin seems in line with Taruya et al.

% Their fig. 1 confirms that $\kappa_min$ decreases with angular scale as one would expect, but that this relationship has started to plateau at 10 arcmin. One would therefore expect $\kappa_min$ to be around 0.01 - 0.02     

%This could be due to spatial correlation of the shape noise in the data, which we have not modeled when adding shape noise to the simulation. Or it could be that there is some non-linearity, which is what the lognormal model is designed to capture, in the data which is not present in the simulation.

%We therefore use the theoretically derived $\kappa_0=0.039$ described in section \ref{sec:LN_theory:WL}. 
The $\kappa_{\rm{WL}}$ distribution appears quite Gaussian at all scales due to the Gaussian shape noise, the distribution of which has a width of 70-90\% of the width of the $\kappa_{\rm{WL}}$ distribution. Despite this low signal to noise, as in the case of simulated $\kappa_{\rm{WL}}$, we find that the lognormal model with Gaussian shape noise (black line) provides a better fit than the simple Gaussian model (magenta line) at small scales. At 10\arcmin\ the lognormal model has $\chi^2 = 1.11$ and the Gaussian 1.84, corresponding to p-values of 0.35 for the lognormal model (i.e. within one $\sigma$) and 0.07 for the Gaussian model. 
%There is therefore significant evidence that the lognormal model, with shape noise accounted for, is a better model of the weak lensing convergence distribution than a simple Gaussian at 10 arcmin. 
At 15\arcmin\ the advantage of the lognormal model over the Gaussian is clear, with best-fit $\chi^2$/DOF of 1.01 and 2.13 respectively. At scales larger than this the Gaussian model provides a good fit with best-fit $\chi^2$/DOF of 1.09, 1.06 and 1.14 at 20, 30 and 40\arcmin. The lognormal model is over-fitting the data at these scales, with $\chi^2$/DOF of 0.46, 0.57 and 0.66 at the same scales, so the Gaussian model is sufficient in this regime.

\subsection{Joint galaxy-convergence distribution}
\label{sec:Joint_DES}

\begin{figure}
		\centering
	\includegraphics[scale=0.36]{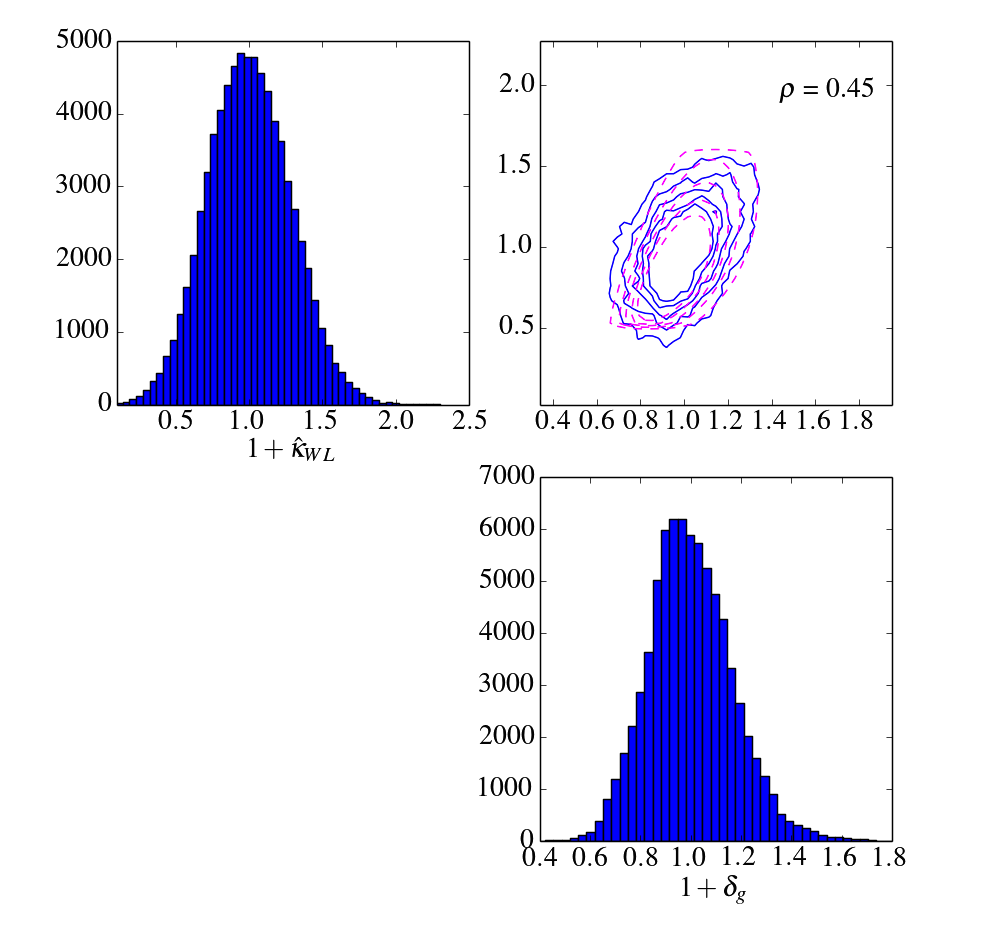}				%\includegraphics[scale=0.42]{DES_scatter_g_0p2z0p5_K_0p3z1p3_mf0p8_cov100_v7_TPZ_Kgal_r30.png}
	\caption{Joint distribution of weak lensing convergence and galaxy density contrast for DES at smoothing scale 15\arcmin. Upper right panel: Fit of bivariate lognormal to DES SV data. Contours for the data are given by the solid blue lines, with dashed magenta contours for the fit. Also shown are the individual 1D PDFs for $1+\delta_g$  and $1+\hat{\kappa}_{WL}$. DES Benchmark galaxies are used, selecting the redshift range $0.1 < z < 0.5$ and WL sources from the imshape catalogue are used over the range $0.6 < z < 1.3$. All redshifts are best-fits from the Skynet pipeline. PDFs are calculated via the CiC method with cells of radius 15\arcmin. This joint distribution has a Pearson correlation coefficient of $r = 0.45$. 
    %LOWER PANEL: The same but now showing the joint and individual CiC distributions for $1+\hat{\kappa}_{gal}$ and $\hat{\kappa}_{WL}$, with correlation coefficient 0.46.
   }
\label{fig:2Dscatter_DES}
\end{figure}

The joint distribution of DES galaxy density contrast and weak lensing convergence data at an angular scale of 15\arcmin\ is shown in the top right panel of fig. \ref{fig:2Dscatter_DES}. The data are shown by the blue contours, and the bivariate fit is shown by the dashed magenta contours. The individual 1D PDFs for $1+\delta_g$  and $1+{\kappa/\kappa_0}$ are also shown. 

Before we account for shot noise in the galaxies and shape noise in the convergence, the galaxy counts and $\kappa_{\rm{WL}}$ have a Pearson correlation coefficient of 0.45. This is in line with what we see in the MICE simulations once DES-like shape noise is added (bottom row of Fig. \ref{fig:2Dscatter_MICE}). 
% The lower panel of this figure shows the same but for the joint and individual CiC distributions for $1+{\kappa_{gal}/\kappa_0}$ and $1+{\kappa/\kappa_0}$.  , $\kappa_0$ is fitted for $\kappa_{gal}$ and 
%As in the 1D case the theoretical minimum convergence, $\kappa_0=0.039$ is used for $\kappa$. 

Once we account for these sources of noise, the correlation coefficient is 0.82, again in line with the noise-free MICE simulations, where the Pearson correlation coefficient was 0.83 (top row of Fig. \ref{fig:2Dscatter_MICE}).

% As expected the re-weighted galaxy map, $\kappa_{gal}$, is better correlated with the weak lensing convergence distribution. ; $\kappa_{gal}$ and $\kappa_{WL}$  have correlation coefficient 0.51.

% We expect these joint distribution to look different to those of the simulations as shape noise in $\kappa$ (which isn't present in the simulations) as well systematics such as photo-z errors will smear out the correlation. These effects could be modelled in the simulations to test whether they fully account for the difference we see, but we leave this to future work.
% [**FURTHER COMMENT?**]

% \begin{figure}
% 		\centering 
% 		\includegraphics[scale=0.4]{DES_2Dfit_g_0p2z0p5_K_0p6z1p3_r15_cond_prob.png}
% %         		\includegraphics[scale=0.4]{DES_2Dfit_g_0p2z0p5_K_0p3z1p3_v7_TPZ_Kgal_r30.png}
% 		\caption{Fit of bivariate lognormal to DES SV data at a smoothing scale of 15 arcmin. Contours for the data are given by the solid blue lines, with dashed magenta contours for the fit. Best fit values of the three free parameters are given.}
%         \label{fig:2D_LN_fit_DES}
% \end{figure}

\subsection{Comparison of Second Moments}
\label{sec:consistency_second_moments}

\begin{figure}
		\centering
        \includegraphics[scale=0.35]{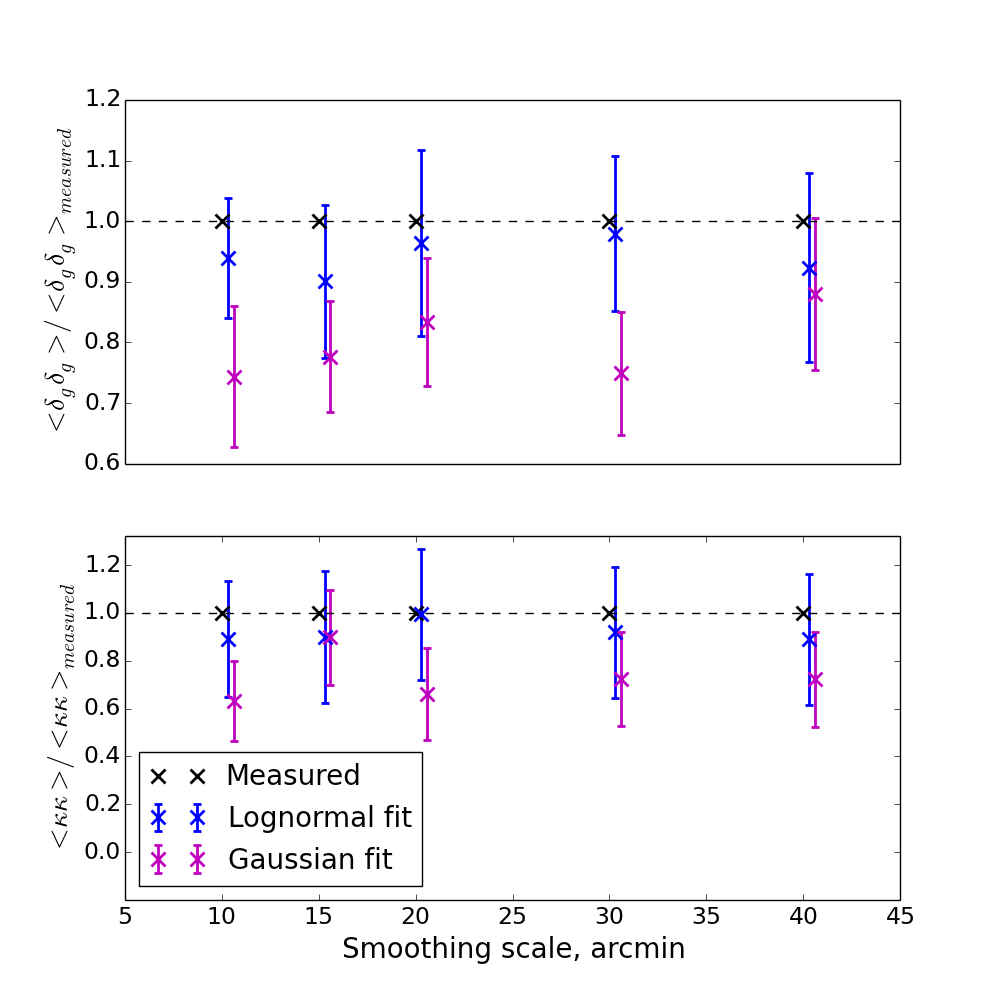}
		\caption{Same as figure \ref{fig:mom2MICE}, but for DES galaxies (upper panel) and convergence (lower panel).}%. as derived directly from CiC versus under lognormal modelling of the CiC PDF.}
        \label{fig:mom2DES}
\end{figure}

\begin{table*}
\centering
\begin{tabular}{|c|c|c|c|}
\hline
% $r$, arcmin & $\langle \delta\delta \\rangle$ & $\langle \kappa \kappa \\rangle$ & $\langle\ kappa_{noise} \kappa_{noise} \\rangle$ \\
$r$, arcmin & $\langle \delta_g\delta_g\rangle$ & $\langle \kappa\kappa\rangle$ & $\langle\ \kappa\kappa\rangle_{SN}$   \\
\hline
10 & $(3.70 \pm 0.22)\times 10^{-2}$    & $(2.52 \pm 0.41)\times 10^{-5}$     & $(1.00 \pm 0.03)\times 10^{-4}$  \\
15 & $(2.76 \pm 0.20)\times 10^{-2}$    & $(1.69 \pm 0.32)\times 10^{-5}$     & $(4.68 \pm 0.18)\times 10^{-5}$  \\
20 & $(2.26 \pm 0.18)\times 10^{-2}$    & $(1.39 \pm 0.25)\times 10^{-5}$     & $(2.60 \pm 0.16)\times 10^{-5}$  \\
30 & $(1.65 \pm 0.14)\times 10^{-2}$    & $(9.84 \pm 1.66)\times 10^{-6}$     & $(1.18 \pm 0.06)\times 10^{-5}$  \\
40 & $(1.38 \pm 0.16)\times 10^{-2}$    & $(8.40 \pm 1.30)\times 10^{-6}$     & $(6.75 \pm 0.44)\times 10^{-6}$  \\
\hline
\end{tabular}
\caption{Second moments of DES galaxy density contrast and weak lensing convergence, as calculated by CiC, for different cell radii. Shot and shape noise have been accounted for, and these are the de-noised moments. The final column gives our estimate the shape noise of the weak lensing convergence. This is derived from the 100 realisations of the $\kappa_{\rm{WL}}$ map with randomised shears, which we find to agree with the second moment of the $\kappa_{\rm{WL}}$ B-mode within 2\%.}
\label{table:ellcuts}
\end{table*}

% & $\langle\kappa\kappa\rangle_{noisy}$
% & $(1.24 \pm 0.03)\times 10^{-4}$
% & $(6.36 \pm 0.26)\times 10^{-5}$
% & $(3.99 \pm 0.21)\times 10^{-5}$
% & $(2.17 \pm 0.15)\times 10^{-5}$
% & $(1.51 \pm 0.12)\times 10^{-5}$

In this section we check the validity of the lognormal model by comparing second moments derived from the log-normal assumption with those measured directly from the data. 
%We calculate second moments directly from the galaxy and convergence fields as well as the joint second moment of the two fields.
%In \ref{sec:moments_formalism} we summarise the formalism for the calculation of the moments of a distribution using the CiC method, and in section \ref{sec:moments2_results} we compare the second moments from the 1-point and bivariate log-normal models with measurements and with theoretical predictions.

% \subsection{}
% \label{sec:moments2_results}

The variance of the DES galaxy PDF is shown in the first panel of Fig. \ref{fig:mom2DES}. Blue data points show the ratio of the variance $\left< \delta_g^{2} \right>$ from our fits to the 1D lognormal distribution to that calculated directly from the CiC PDF. Errors on $\left< \kappa^{2} \right>$ directly from CiC are produced by jackknife sampling; errors on $\left< \kappa^{2} \right>$ derived from the lognormal fit are from the 1$\sigma$ width of the likelihood of the lognormal width. The second panels shows the same for DES $\kappa_{\rm{WL}}$. 
The lognormal model with appropriate noise contribution gives an estimate of the variance that is consistent with that calculated directly from the CiC, for both galaxies and $\kappa_{\rm{WL}}$, at all scales from 10 - 40\arcmin. 

For the galaxy density contrast distribution, the Gaussian model provides a less accurate estimate of the variance calculated directly from the CiC at all scales. 
%Agreement of these variance estimates improves with scale with the exception on 30\arcmin, where it can be seen from fig. \ref{fig:smoothing} (top row) that the best-fit Gaussian PDF is narrower than the CiC distribution. 
For the convergence distribution the Gaussian model again gives variance estimates less accurate than the lognormal model at all scales. 
%The agreement at 15\arcmin is very good but while the widths of the best-fit lognormal and Gaussian PDFs are very similar the lognormal provides a better fit (see fig. \ref{fig:smoothing}). 

For both galaxies and weak lensing convergence, the Gaussian and lognormal approaches underestimate the variance as compared to measuring it directly from the CiC. This is because in constructing the CiC PDF to which we fit the lognormal model, we bin the cell counts. We account for noise via singular value decomposition, and one of the things this removes is contributions to the fit from the outermost bins, which have very few cell counts. This makes the effective distribution narrower, with lower second moment, than if these noisy data points were included. In calculating the variance directly from the CiC (as described in Appendix \ref{sec:Ap_mom}) this binning is not necessary and all cells, including those with the most extreme values of $\delta_g$ or $\kappa_{\rm{WL}}$, are included in the calculation, resulting in a larger variance in $\delta_g$ or $\kappa_{\rm{WL}}$. This effect is less stark in the MICE simulations where there are a greater number of galaxies than in the DES data, so fewer bins are discarded due to low counts of cells. This underestimation of the variance, however, is not significant within the errors.

\section{Discussion}
\label{sec:conc}

We have tested the lognormality of the DES galaxy density contrast and weak lensing convergence PDFs at angular scales of 10 - 40\arcmin\ (corresponding to physical scales of 3 - 10 Mpc at median redshift $z=0.3$). In the context of this work, estimating the CiC PDF is a way of quantifying the non-linear growth of mass and galaxy fluctuations, as well as the visual impression of comparing the $\kappa_{\rm{WL}}$ mass maps with the galaxy distribution on the same patch of the sky. It is also a test of systematics. 
\vspace{5mm}
Our main findings are as follows:

\begin{itemize}
  
 \item In agreement with many earlier papers we find that the 1D DES galaxy PDF is well fitted by a lognormal model, taking into account Poisson shot noise, with best-fit $\chi^2$/DOF$ = 1.28$ vs. 6.55 for a Gaussian model at a scale of 10\arcmin.

\item In modelling the weak lensing convergence distribution it is important to account for shape noise since the width of this noise is a significant fraction (70-90\%) of the width of the $\kappa_{\rm{WL}}$ signal. We find that the shape noise estimate derived from the 100 realisations of DES $\kappa_{\rm{WL}}$ in which the shears have been randomised agrees with that of the $\kappa_B$ mode within 2\% at all scales from 10 to 40\arcmin, and that the distribution of the shape noise can be well modeled by a Gaussian PDF. This allows us to model the $\kappa_{\rm{WL}}$ distribution with a lognormal convolved with Gaussian PDF. In future work it would be interesting to investigate the spatial correlation of this noise.  

\item The convergence field is not expected to be exactly lognormal even if the mass density contrast field is, as it is a weighted projection of the mass density field along the line of sight. We find however, in agreement with previous work on simulations, that the 1D $\kappa_{\rm{WL}}$ PDF is well fitted by a lognormal model, taking into account shape noise. This is the first such measurement on data.  The best-fit $\chi^2$/DOF for the lognormal model is 1.11, compared to 1.84 for a Gaussian model, corresponding to p-values of 0.35 (i.e. within one $\sigma$) and 0.07 respectively. At scales above 15\arcmin\ the Gaussian model is a sufficient approximation. 

\item The bivariate ($\kappa_{\rm{WL}}$, $\delta_g$) PDF is also well fitted by a bivariate lognormal. 

\item De-noised second moments derived via the lognormal fit are consistent with variances derived directly from the data up to scales of 40\arcmin, for both the DES galaxy density contrast and weak lensing convergence distributions.

\end{itemize}

%- [ADD COMMENT ABOUT rho, rLN]
% - As  kappa is a sum over a weighted matter overdensity field, this suggests that the underlying matter density field is also characterised by a lognormal, as previously seen in LCDM N-body simulations.

% - Derived parameters from the log normal fits are useful for constraining non-linear and stochastic biasing.

This pilot study could be extended to much larger areas with weak lensing surveys such as the full DES (5000 deg$^2$) survey, LSST (20,000 deg$^2$) and Euclid (15,000 deg$^2$). In this work we have tested the lognormality of the $\kappa_{\rm{WL}}$ PDF; with the higher signal/noise that future surveys will provide it might be possible to deduce from the observed $\kappa_{\rm{WL}}$ PDF whether or not the underlying matter density field is lognormal - essentially inverting equation 4. 

In this work we have used the CiC to probe lognormality, but there is a wealth of information contained within it that could be exploited in future work. The CiC contains the full PDF so as well as the variance, higher order moments such as skewness and kurtosis can also be extracted. 

The method used in this work, of constructing PDFs via CiC and cross correlating them, could be used to extract information on galaxy bias and to derive cosmological parameters. It could also be interesting to repeat this analysis using manipulations of the shear field than other $\kappa_{\rm{WL}}$ that avoid the reconstruction noise due to the Kaiser Squires method. 

%of constructing the weak lensing convergence.
% shear rather
% with
% than $\kappa_{\rm{WL}}$ as this would avoid the reconstruction noise due to the Kaiser Squires method of constructing the weak lensing convergence.

Quantifying $P(\kappa_{\rm{WL}})$ will be important for the emerging field of mass reconstruction using $\kappa_{\rm{WL}}$, since it is required as a prior input for this process. We have demonstrated that a lognormal model is a better choice than a Gaussian model at scales of 10 - 20\arcmin. As well as the improved ability to capture non-linear behaviour versus a Gaussian model, the lognormal model still allows fast production of, for example, simulated realisations of the convergence field for testing, and covariance matrices.

% -  Systematic maps (e.g. stellar occultation, extinction, seeing etc) exhibit very different behaviour than the galaxy and mass maps, suggesting the log-normal fits don’t are not contaminated by these systematics.

% One of the benefits of the CiC approach is that it preserves information about the location on the sky of analysis elements i.e. there is no mixing of spatial locations beyond the limits of a single cell. This is especially important when checking the impact of spatially varying systematics such as stellar occultation, extinction, seeing etc. We conducted an extensive analysis correlating our observable with maps of each observational systematic. No significant correlations were found, suggesting that our analysis is robust to bias by these phenomena.

\appendix

\section{Measurement of moments from counts-in-cells}
\label{sec:Ap_mom}

% The moments of a distribution characterise that distribution. The first and second moments are the mean and variance of the distribution, higher moments like skewness (third moment) and kurtosis (fourth moment) contain information beyond that of a Gaussian statistic like the two point correlation function or the power spectrum. 

In section \ref{sec:MICE} we use the second moment, as calculated via CiC, as a check that the lognormal model accurately recovers the characteristics of the galaxy and $\kappa_{\rm{WL}}$ distributions. In this section we show how these are calculated for the galaxy and weak lensing convergence distributions, including how noise is accounted for. 

Moments of a distribution are easily obtained via the CiC technique, with the $p^{th}$ central moment of the distribution of the number of objects $n$ at angular scale $\theta$ given by:

\begin{equation}\label{eq:moments}
 	m_p(\theta) = \frac{1}  {N(\theta)} \sum_{i=1}^{n_c(\theta)} \left( x_i(\theta) - \bar{x}\right)^p 
\end{equation}
%= \bar{x}^p(\theta)\langle\delta_i^p(\theta) \rangle
%
where $N(\theta)$ is the number of cells of angular size $\theta$ used, $\bar{x}$ is the mean count of observable $x$ in a cell,  and $x_i$ is the count in cell $i$. For the distribution of galaxies $x_i$ is the number of galaxies in a cell, and for the convergence $x_i$ is the average $\kappa$ within a cell.

%  One can continue calculating moments of higher order though, in practice, measurements from data become increasingly noisy. In this paper we calculate moments up to fourth order.

The connected central moments $\mu_p(\theta)$ can be derived using the moment generating function (see §3.2.4 of \citealt{Bernardeau2002} for a derivation and nice diagrammatic representation of the connected moments). 
% Since $\langle\delta_g \rangle = 0$ 
The second connected moment is equal to the second central moment, $\mu_2 = m_2$.
 
% \begin{eqnarray}\label{eq:mu_p}
%  	\mu_2 &=& m_2,\\
% %    	\mu_3 &=& m_3,\\
% %    	\mu_4 &=& m_4 - 3m_2^3.
% \end{eqnarray}

For the galaxy distribution, shot noise can be accounted for by assuming that galaxies form a  Poisson sampling of the underlying matter density field:
  
 \begin{equation}\label{eq:mom}
 	P(\lambda) = \sum_{n=0}^{\infty} \lambda^n P(n). 
\end{equation}
 Taylor expanding this around $\lambda = 1$ gives $\langle n(n-1)...(n-p-1) \rangle = \bar{n}^pk_p$ where $k_p$ is the $p^{th}$ moment of the local density distribution. 
%  The first few of these `de-noised' moments are:
The `de-noised' second moment is:
%  \begin{eqnarray}\label{eq:k_p}
%  	k_2 &=& \mu_2 - \bar{n},\\
%    	k_3 &=& \mu_3 - 3k_2 - \bar{n},\\
%    	k_4 &=& \mu_4 - 7k_3 - 6k_2 - \bar{n}.
% \end{eqnarray}

 \begin{equation}\label{eq:k_p}
 	k_2 = \mu_2 - \bar{n},\\
\end{equation}
where $\bar{n}$ is the mean number of galaxies in a cell. The area averaged correlations are then given by

\begin{equation}\label{eq:w_p}
	\bar{w}_p = \frac{k_p}{\bar{n}_p}, 
 \end{equation}
    
%    and the quantities we will look at are the \textit{hierarchical moments} given by the ratios $S_n \equiv \frac{\bar{w}_p} {\bar{w}_2^{p-1}}$.
%
 For both the MICE and DES galaxy distributions, the second moment as calculated via CiC, with shot noise removed, is given by equations \ref{eq:moments} - \ref{eq:w_p}.

For $\kappa_{WL}$, since there is no need to model shot noise, $k_2 = \mu_2$. The second moment for MICE $\kappa_{WL}$ is then given by equation \ref{eq:w_p}. In the case of DES $\kappa$ we need to remove shape noise. The shape noise in the DES $\kappa_{WL}$ map is estimated from the 100 noise realisations, as discussed in section \ref{data:DES_K}. Following Van Waerbeke et al. 2013, we assume the de-noised second moment of DES $\kappa_{WL}$ is then given by

\begin{equation}\label{w2_kappa}
	\bar{w}_2 = \bar{w}_{2,data} - \bar{w}_{2,noise}
\end{equation}
where $\bar{w}_{2,noise}$ is given by equations \ref{eq:moments} - \ref{eq:w_p} (with $k_2 = \mu_2$), and  $\bar{w}_{2,noise}$ is the mean of the second moments measured via CiC from each of the noise maps. 

The second moment, estimated jointly from two distributions becomes

\begin{eqnarray}\label{eq:joint_moments}
 	m_2(\theta) &=& \frac{1}  {N(\theta)} \sum_{i=1}^{n_c(\theta)} \left( n_{\delta,i}(\theta) - \bar{n}_{\delta}\right)\left( \kappa_i(\theta)- \bar{\kappa} \right) \\
    &=& \bar{n}_{\delta}\bar{n}_{\kappa}(\theta)\langle\delta_i(\theta)\kappa_i(\theta) \rangle
\end{eqnarray}

\section{Systematic Effects}
\label{sec:Ap_syst}

\begin{figure*}
		\centering
		\includegraphics[scale=0.35]{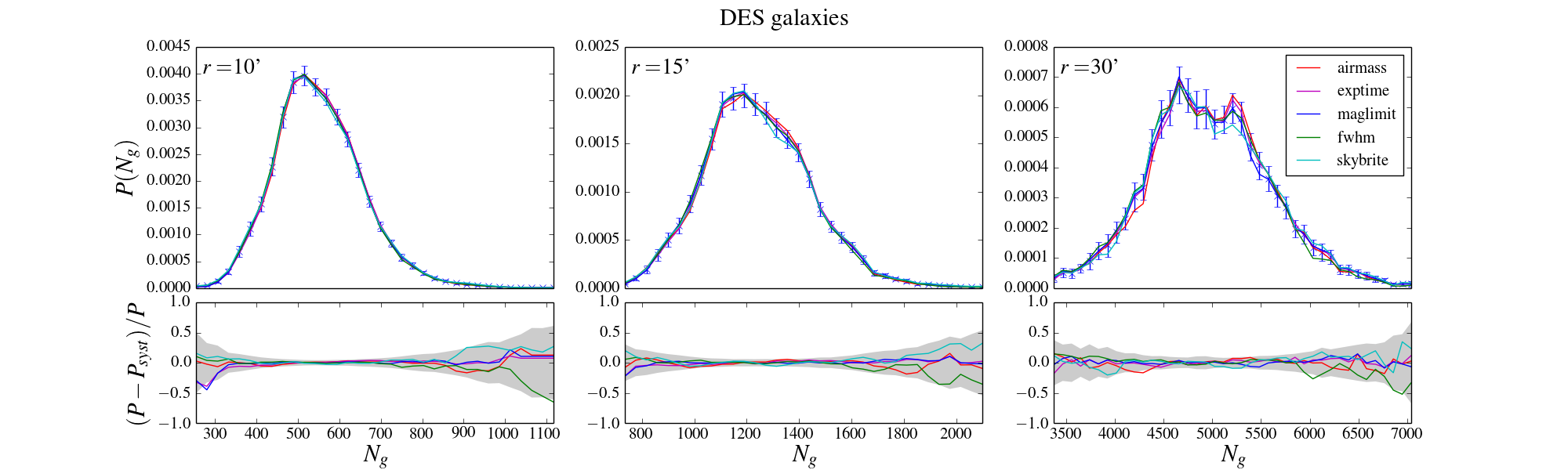}
\includegraphics[scale=0.35]      {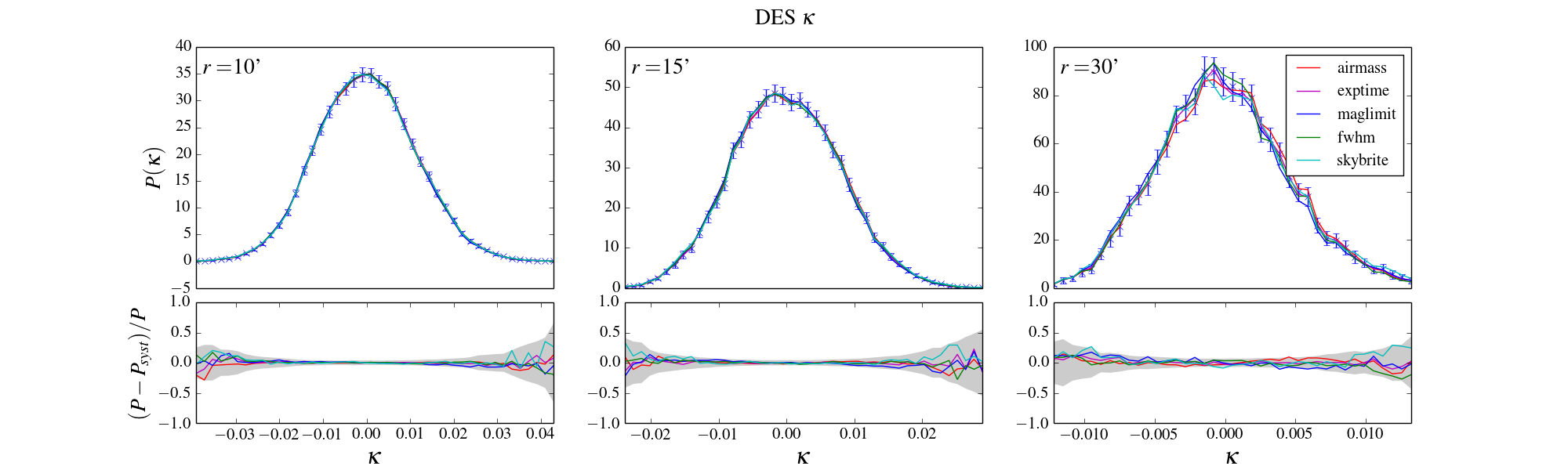}
\caption{\label{fig:syst} \textbf{UPPER ROW:} PDF of DES galaxy number density at cell radii of 10, 15, 30\arcmin\ (blue data points) with jackknife errors. Coloured lines show PDFs with cuts for each systematic effect in turn. Lower panel shows the fractional difference between the full sample and those with systematics cuts, with errors shown by the grey shaded region.  \textbf{LOWER ROW:} Same but for DES $\kappa_{WL}$. }
\end{figure*}

We investigate the potential impact on our results of spatially varying systematics. 
%on the probability distributions of DES galaxies and weak lensing convergence. 
The systematics we consider are varying the amount of air mass dependent on the distance of the observed field from the zenith, exposure time, magnitude limit, atmospheric seeing, and sky brightness. The values of these properties were mapped across the DES-SV area as described in \citet{Leistedt2015}.

We compare PDFs of $\delta_g$ and $\kappa_{\rm{WL}}$ for the full samples used in this work versus when the areas worst-affected by these systematics are removed. We produce PDFs with the 20\% of worst-affected pixels masked, for each systematic in turn. 

Figure \ref{fig:syst} shows the resulting distributions. The top row shows DES galaxy number density at cell radii of 10, 15, 30\arcmin\ (blue data points) with jackknife errors. Coloured lines show PDFs with cuts for each systematic effect in turn. The lower panel shows the fractional difference between the full sample and those with systematics cuts, with errors shown by the grey shaded region. The same for DES $\kappa_{\rm{WL}}$ is shown in the bottom row. 

Here we can see that the PDFs of the cut data are broadly consistent with that of the full data, given the jackknife errors. For DES galaxies, for each systematic effect at least 95\% of the bin heights after the cuts are made fall within the jackknife errors of the original distribution, and all are within 1.5 sigma of the original distribution. For DES $\kappa_{\rm{WL}}$, at least 93\% of the new bin heights fall within the jackknife errors of the original distribution. All are within 1.9 sigma of the original data points. We can see that the effect of the systematics on the distribution increases with scale. Importantly  PDFs of the cut $\kappa_{\rm{WL}}$ data at scales below 20\arcmin, which is where we detect lognormality of $\kappa_{\rm{WL}}$, are completely consistent with the original distributions, i.e. all of the new bin heights fall within the errors of the original distribution. This simple test is reassuring and indicates that our lognormal fits to the DES $\delta_g$ and $\kappa_{\rm{WL}}$ distributions are not likely to be affected by these systematic effects.

\section{Tests of Sampling Methods}
\label{sec:Ap_sampling}

\begin{figure}
		\centering
		\includegraphics[scale=0.35]{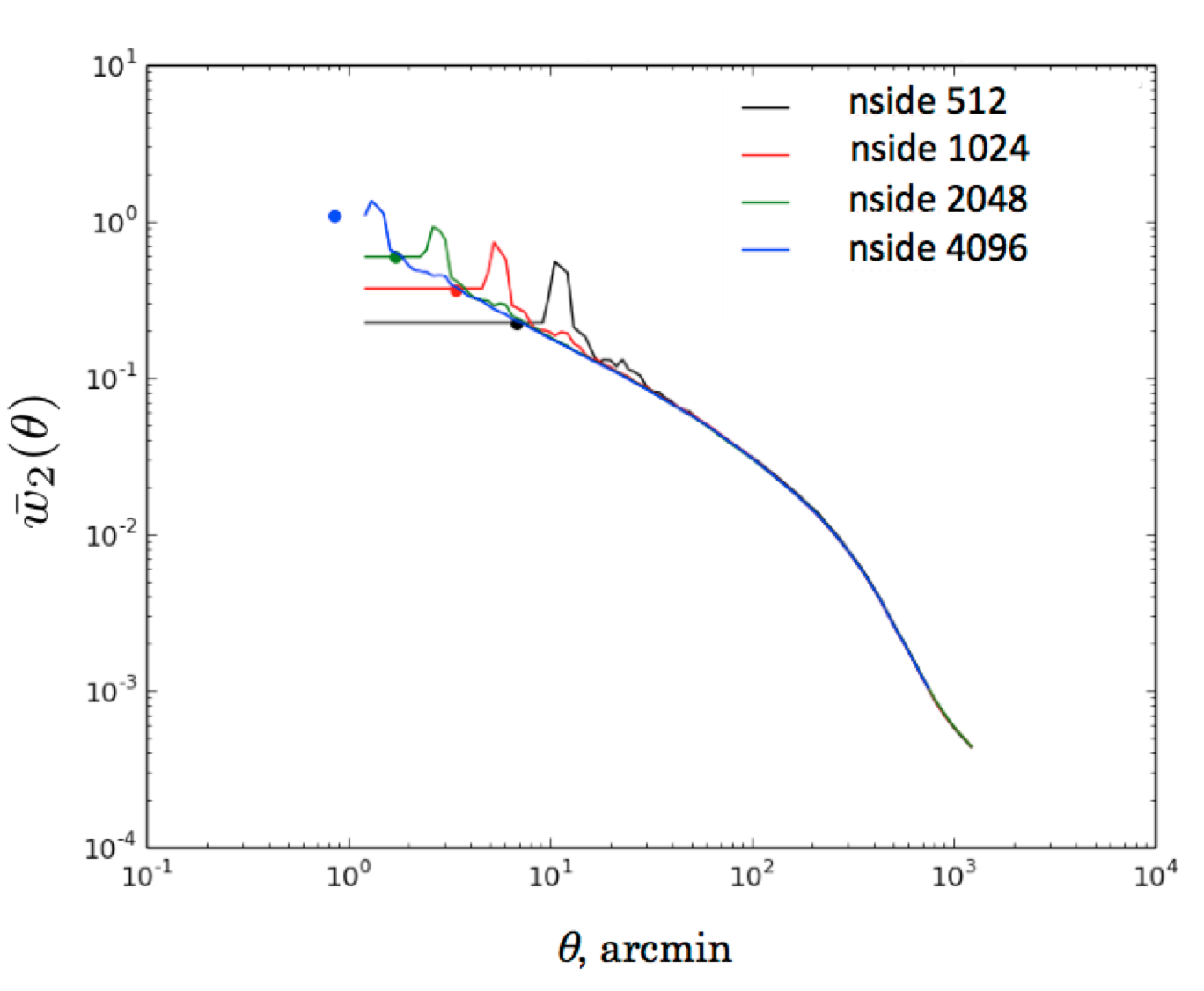}
		\caption{\label{fig:pixel_res} Second moment of MICE galaxy density contrast distribution as a function of scale as calculated via CiC, using underlying \texttt{HEALPix} maps with different resolutions. The \texttt{HEALPix} maps have nside (see main text) 512, 1024, 2048, 4096 corresponding to pixels of sizes shown by the solid circles. At scales approaching the pixel size edge effects are visible. }
\end{figure}

Our CiC analysis has made particular choices for cell size and distribution when accounting for the mask and creating the underlying \texttt{HEALPix} maps. In this Appendix we test each of these assumptions and demonstrate that the conclusions of our analysis are robust to our methodological choices.

\texttt{HEALPix} tessellations are made up of pixels with equal area, but not equal shape. Using circles that encompass several pixels will reduce the effect of the different pixel shapes, more so the larger the circles relative to the pixels. To check that the effect of varying shapes is effectively mitigated in this way we measure the area averaged 2-point correlation $\bar{w}_2(\theta)$ for DES galaxies at different \texttt{HEALPix} resolutions (512, 1024, 2048, and 4096). Figure \ref{fig:pixel_res} shows that when the cells size is close to the pixel size the correlation function is not smooth due to the effects of pixel shape. Once the cells are several times larger than the average pixel separation the correlation function becomes smooth, and the correlation functions based on the different \texttt{HEALPix} resolutions converge. This confirms that the method of using circular cells several times larger than the pixel resolution does not suffer from the effects of pixel shape, and that the underlying pixel resolution is not important as long as the minimum cell size considered is sufficiently large.

%In the preceeding analyses we have used a minimum cell size of 10 times the pixel resolution, here we test the robustness of this. We measure the variance of the DES galaxy CiC distribution at a scale of 35 arcmins at HEALPix resolution $nisde$ 1024, so that the cells are 10 times larger than the pixel size, and again at a finer resolution of $nside$ 4096. 

% As described in section 4, at each cell radius circles are thrown down in random positions and accpeted of 80\% of their area is unmasked. Here we test the sensitivity of measurements of the Variance of the DEs galaxy CiC PDF this by measuring it for acceptence of 50\% and 90\% [figures to be added.]

% An alternative to using circular cells larger than the pixel size is to use the HEALPix pixels themselves as the 'cells'. This means that there is no overlap of cells and the all areas of the survey have equal coverage, but it introduces edge effects due to the slightly different shapes of the pixels, and also limits the scales one can probe to the available HEALPix resolutions. Here we repeat the measurement of vaiance of DES galaxy CiC distribution with this method and compare it to the results obtained with circular cells [figure to be added].

% \section{Tests of Sampling Methods}
% \label{sec:ApB}

\begin{figure*}
		\centering
		\includegraphics[scale=0.4]{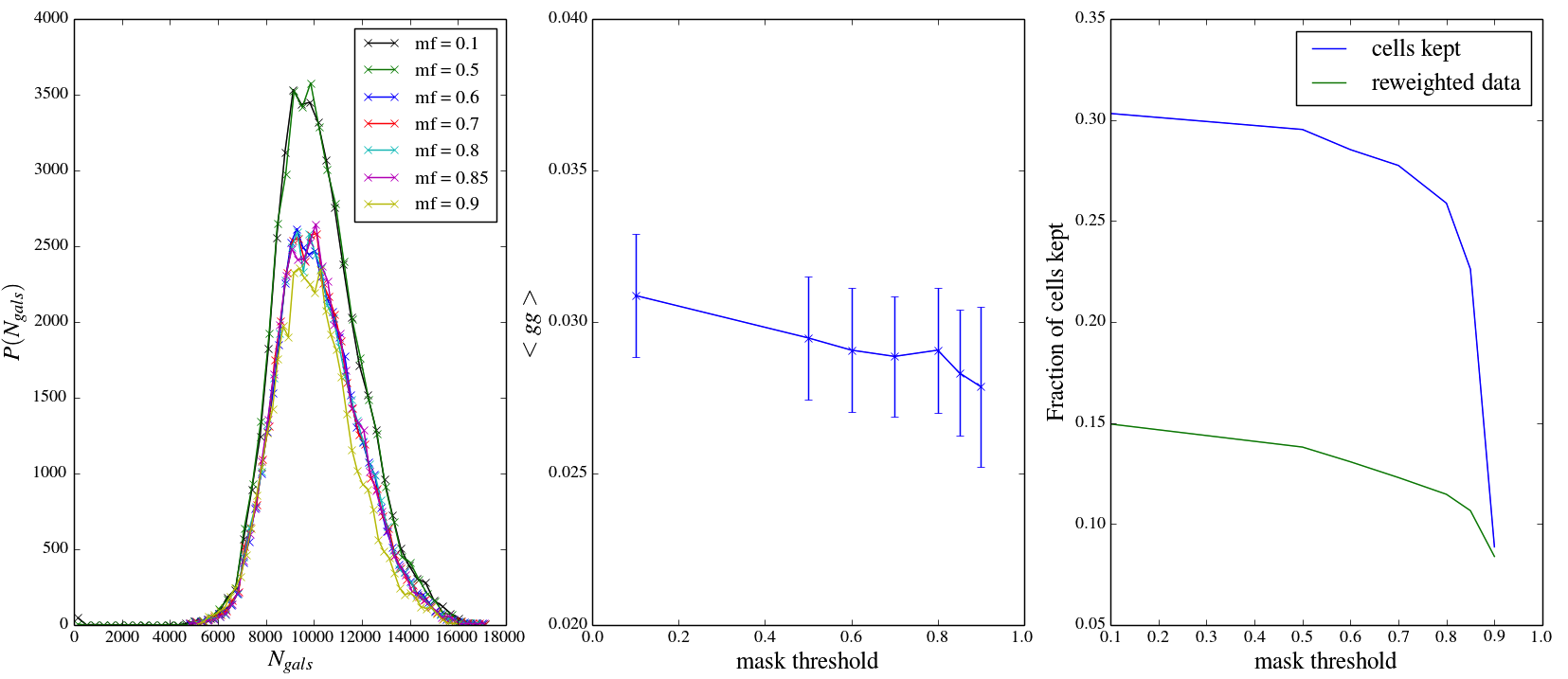}
		\caption{\label{fig:mf} Left panel: Effect of the mask threshold (fraction of a cell that must be unmasked in order to be included in the analysis) on the resulting probability distribution of MICE galaxies. Middle Panel: Effect of this threshold on the second moment. Right panel: The fraction of the cells randomly thrown that are kept in the analysis, and, of those kept, the fraction of the data that comes from re-weighting these cells.}
\end{figure*}

The other sampling assumption we test is the threshold at which we decide to discard a randomly positioned cell because too much of it is masked. If unmasked fraction of a randomly positioned cell  is less than $f$, the counts are up-weighted to make it the equivalent of a whole cell. Here we explain the choice of $f = 0.8$ used in this work. 

The first panel of figure \ref{fig:mf} shows the PDF of MICE galaxy counts for which the $f$ takes different values. For values of $f > 0.5$ there is not much difference in the histograms by eye. The middle panel shows the variances of these different distributions, with errors produced by jackknife sampling. We find that the effect on the variance of changing $f$ is not significant within the errors. 

If we chose a very high threshold, such as requiring 90\% of a cell to be unmasked in order for it to be used, we would throw away a lot of cells landing near the edge of the survey and give greater statistical weight to areas away from the edges. If we set $f$ too low, so that cells with a large fraction of their area masked are kept, we will end up re-weighting a lot the data nearer the edges. So we would like to strike a balance between these two effects. From the right panel of figure \ref{fig:mf} we can see than $f= 0.8$ is the highest value that can be allowed before the number of cells we discard drops off significantly, and that at this value the fraction of data re-weighted is not too high (around 10\%). Hence $f=0.8$ seems to be a sensible choice.

% \begin{figure}
% 		\centering
% 		\includegraphics[scale=0.4]{placeholder_cic_1D_pdf.png}
% 		\caption{\label{fig:z_kernels} 1D PDFs for different CiC procedures - check for systematic effects.}
% \end{figure}

\section*{Acknowledgements}
The authors would like to thank Ludovic Van Waerbeke for extremely useful exchanges on the formalism for the calculation of the convergence second moment.

We are grateful for the extraordinary contributions of our CTIO colleagues and the DECam Construction, Commissioning and Science Verification
teams in achieving the excellent instrument and telescope conditions that have made this work possible.  The success of this project also 
relies critically on the expertise and dedication of the DES Data Management group.

Funding for the DES Projects has been provided by the U.S. Department of Energy, the U.S. National Science Foundation, the Ministry of Science and Education of Spain, the Science and Technology Facilities Council of the United Kingdom, the Higher Education Funding Council for England, the National Center for Supercomputing Applications at the University of Illinois at Urbana-Champaign, the Kavli Institute of Cosmological Physics at the University of Chicago, the Center for Cosmology and Astro-Particle Physics at the Ohio State University, the Mitchell Institute for Fundamental Physics and Astronomy at Texas A\&M University, Financiadora de Estudos e Projetos, Funda{\c c}{\~a}o Carlos Chagas Filho de Amparo {\`a} Pesquisa do Estado do Rio de Janeiro, Conselho Nacional de Desenvolvimento Cient{\'i}fico e Tecnol{\'o}gico and the Minist{\'e}rio da Ci{\^e}ncia, Tecnologia e Inova{\c c}{\~a}o, the Deutsche Forschungsgemeinschaft and the Collaborating Institutions in the Dark Energy Survey. 

The Collaborating Institutions are Argonne National Laboratory, the University of California at Santa Cruz, the University of Cambridge, Centro de Investigaciones Energ{\'e}ticas, Medioambientales y Tecnol{\'o}gicas-Madrid, the University of Chicago, University College London, the DES-Brazil Consortium, the University of Edinburgh, the Eidgen{\"o}ssische Technische Hochschule (ETH) Z{\"u}rich, Fermi National Accelerator Laboratory, the University of Illinois at Urbana-Champaign, the Institut de Ci{\`e}ncies de l'Espai (IEEC/CSIC), the Institut de F{\'i}sica d'Altes Energies, Lawrence Berkeley National Laboratory, the Ludwig-Maximilians Universit{\"a}t M{\"u}nchen and the associated Excellence Cluster Universe, the University of Michigan, the National Optical Astronomy Observatory, the University of Nottingham, The Ohio State University, the University of Pennsylvania, the University of Portsmouth, SLAC National Accelerator Laboratory, Stanford University, the University of Sussex, Texas A\&M University, and the OzDES Membership Consortium.

The DES data management system is supported by the National Science Foundation under Grant Number AST-1138766.
The DES participants from Spanish institutions are partially supported by MINECO under grants AYA2012-39559, ESP2013-48274, FPA2013-47986, and Centro de Excelencia Severo Ochoa SEV-2012-0234.
Research leading to these results has received funding from the European Research Council under the European Union’s Seventh Framework Programme (FP7/2007-2013) including ERC grant agreements 
 240672, 291329, and 306478.

This paper has gone through internal review by the DES collaboration.

LC thanks the Perren Fund for a studentship. 

OL, DK and  MM  acknowledge support from a European Research Council Advanced Grant FP7/291329.

\bibliographystyle{mnras}
\bibliography{cic}

%\bsp

% \appendix
% \section{Author Affiliations}
% \label{sec:affiliations}
% \input{affil.tex}

~
\newline

~
\newline
$[1]$ Astrophysics Group, Department of Physics and Astronomy, University College London, 132 Hampstead Road, London, NW1 2PS, United Kingdom \\ 
$[2]$ Department of Physics and Electronics, Rhodes University, PO Box 94, Grahamstown, 6140, South Africa \\
$[3]$ Department of Physics, ETH Zurich, Wolfgang-Pauli- Strasse 16, CH-8093 Zurich, Switzerland \\ 
$[4]$ Institute of Cosmology \& Gravitation, University of Portsmouth, Portsmouth, PO1 3FX, UK \\
$[5]$Institut de Ci\'encies de l'Espai (ICE, IEEC/CSIC), E-08193 Bellaterra (Barcelona), Spain \\
$[6]$ Istituto Nazionale di Astro sica - Osservatorio Astronomico di Brera, Via E. Bianchi 46, 23807 Merate, Italy\\
$[7]$ Department of Physics and Astronomy, University of Pennsylvania, Philadelphia, PA 19104, USA \\
$[8]$ Cerro Tololo Inter-American Observatory, National Optical Astronomy Observatory, Casilla 603, La Serena, Chile \\
$[9]$ Fermi National Accelerator Laboratory, P. O. Box 500, Batavia, IL 60510, USA\\
$[10]$ Department of Astrophysical Sciences, Princeton University, Peyton Hall, Princeton, NJ 08544, USA\\
$[11]$ CNRS, UMR 7095, Institut d'Astrophysique de Paris, F-75014, Paris, France\\
$[12]$ Sorbonne Universit\'es, UPMC Univ Paris 06, UMR 7095, Institut d'Astrophysique de Paris, F-75014, Paris, France\\
$[13]$ Carnegie Observatories, 813 Santa Barbara St., Pasadena, CA 91101, USA\\
$[14]$ Kavli Institute for Particle Astrophysics \& Cosmology, P. O. Box 2450, Stanford University, Stanford, CA 94305, US\\
$[15]$ SLAC National Accelerator Laboratory, Menlo Park, CA 94025, USA\\
$[16]$ Laborat\'orio Interinstitucional de e-Astronomia - LIneA, Rua Gal. Jos\'e Cristino 77, Rio de Janeiro, RJ - 20921-400, Brazil\\
$[17]$ Observat\'orio Nacional, Rua Gal. Jos\'e Cristino 77, Rio de Janeiro, RJ - 20921-400, Brazil\\
$[18]$ Department of Astronomy, University of Illinois, 1002 W. Green Street, Urbana, IL 61801, US\\
$[19]$ National Center for Supercomputing Applications, 1205 West Clark St., Urbana, IL 61801, USA\\
$[20]$ School of Physics and Astronomy, University of Southampton,  Southampton, SO17 1BJ, UK\\
$[21]$ Excellence Cluster Universe, Boltzmannstr.\ 2, 85748 Garching, Germany\\
$[22]$ Faculty of Physics, Ludwig-Maximilians-Universit\"at, Scheinerstr. 1, 81679 Munich, Germany\\
$[23]$ Jet Propulsion Laboratory, California Institute of Technology, 4800 Oak Grove Dr., Pasadena, CA 91109, USA\\
$[24]$ Department of Astronomy, University of Michigan, Ann Arbor, MI 48109, USA\\
$[25]$ Department of Physics, University of Michigan, Ann Arbor, MI 48109, USA\\
$[26]$ Kavli Institute for Cosmological Physics, University of Chicago, Chicago, IL 60637, USA\\
$[27]$ Center for Cosmology and Astro-Particle Physics, The Ohio State University, Columbus, OH 43210, USA\\
$[28]$ Department of Physics, The Ohio State University, Columbus, OH 43210, USA\\
$[29]$ Australian Astronomical Observatory, North Ryde, NSW 2113, Australia\\
$[30]$ Departamento de F\'{\i}sica Matem\'atica,  Instituto de F\'{\i}sica, Universidade de S\~ao Paulo,  CP 66318, CEP 05314-970, S\~ao Paulo, SP,  Brazil\\
$[31]$ Instituci\'o Catalana de Recerca i Estudis Avan\c{c}ats, E-08010 Barcelona, Spain\\
$[32]$ Institut de F\'{\i}sica d'Altes Energies (IFAE), The Barcelona Institute of Science and Technology, Campus UAB, 08193 Bellaterra (Barcelona) Spain\\
$[33]$ Department of Physics and Astronomy, Pevensey Building, University of Sussex, Brighton, BN1 9QH, UK\\
$[34]$ Centro de Investigaciones Energ\'eticas, Medioambientales y Tecnol\'ogicas (CIEMAT), Madrid, Spain\\
$[35]$  ICTP South American Institute for Fundamental Research\\Instituto de F\'{\i}sica Te\'orica, Universidade Estadual Paulista, S\~ao Paulo, Brazil\\

\label{lastpage}

\end{document}